# A comprehensive study of the physical properties of $Nb_2P_5$ via *ab-initio* technique

M. I. Naher, S. H. Naqib*

*Department of Physics, University of Rajshahi, Rajshahi 6205, Bangladesh*

*Corresponding author email: salehnaqib@yahoo.com

## Abstract

Binary metallic phosphide, $Nb_2P_5$, belongs to technologically important class of materials. Quite surprisingly, a large number of physical properties of $Nb_2P_5$, including elastic properties and their anisotropy, acoustic, electronic (DOS, charge density distribution, electron density difference), thermo-physical, bonding characteristics, and optical properties have not been investigated at all. In the present work we have explored all these properties in details for the first time employing density functional theory based first-principles method. $Nb_2P_5$ is found to be a mechanically stable, elastically anisotropic compound with weak brittle character. The bondings among the atoms are dominated by covalent and ionic contributions with small signature of metallic feature. The compound possesses high level of machinability. $Nb_2P_5$ is a moderately hard compound. The band structure calculations reveal metallic conduction with a large electronic density of states at the Fermi level. Calculated values of different thermal properties indicate that $Nb_2P_5$ has the potential to be used as a thermal barrier coating material. The energy dependent optical parameters show close agreement with the underlying electronic band structure. The optical absorption and reflectivity spectra and the static index of refraction of $Nb_2P_5$ show that the compound holds promise to be used in optoelectronic device sector. Unlike notable anisotropy in elastic and mechanical properties, the optical parameters are found to be almost isotropic.

## 1. Introduction

Among all the elements in the periodic table, phosphorus (P) is one of the most active chemically. Phosphorous can exist as a standalone anion as well as in a network of polyanions with covalent P−P bonds that can usually be found in different forms of metal phosphides with different stoichiometries ranging from metal-rich phosphides (e.g., $T_3P$, where T denotes metallic element), to phosphorous-rich polyphosphides (such as $TP_4$, where T = Cr, Mn, and Fe) [1-3]. In recent decades a large family of metal phosphides has been drawing continuous interest of physics, chemistry and materials science research community because of its earth-abundance as well as their diverse and sometimes exotic physical and chemical properties. It has been observed that various physical properties of metal phosphides, from metallic behavior to diamagnetic semiconductor or insulator are closely related to the bonding states between metals and phosphorus [4]. Some of the metal phosphides do not exhibit superconductivity at normal pressure, but they could become superconductors through chemical doping or by applying external pressure. For example, Rh-doped RuP ($Ru_{0.55}Rh_{0.45}P$) becomes superconducting at $T_c$ = 3.7 K [5] and the black phosphorus exhibits superconductivity at $T_c$ up to 9.5 K when 32 GPa pressure is applied [6]. However, superconductivity in binary metal phosphides, with chemical compositions of $T_xP_y$, are the minorities among the various families of superconductors. For example, $Mo_3P$, $Mo_8P_5$, and $Mo_4P_3$ are superconductors at transition temperatures ($T_c$) of 7 K, 5.8 K and 3.0 K, respectively [7].

It is well known that P has a very high vapor pressure; it is therefore, difficult to prepare metal phosphides by conventional methods, especially those phosphorus rich metal phosphides. The high-pressure technique has been proven to be efficient in synthesizing ternary metal rich compounds, such as NbPS ($T_c$ = 12 K) [8] etc. $ZrRu_4P_2$ synthesized at a pressure ~4 GPa and 850 ºC with a tetragonal $ZrFe_4Si_2$-type structure (P42/mnm) became superconducting at ~11 K [9]. The $Fe_2P$-type MoNiP and $Co_2P$-type MoRuP synthesized at ~4 GPa and 1600 ºC have relatively high $T_c$ of about 15.5 K [10, 11]. All these findings indicate that metal phosphides are promising materials to explore superconductivity further.

On the other hand, it has recently been found that the metal phosphides show superconductivity with nontrivial topological band structure. For instance, WP was reported first as a superconductor ($T_c$ ~0.8 K) at ambient pressure [12] and also predicted to be a topological high symmetry line semimetal when the spin-orbit coupling (SOC) was considered [13]. Furthermore, triply degenerate fermions in MoP, crystallizing in the WC-type structure, were predicated theoretically and later experimentally verified [14]. At pressures above 30 GPa, the superconductivity and nontrivial topological band structure in MoP are expected to coexist according to a density functional theory (DFT) calculations [15]. However, it is still not established whether there is a correlation between the superconductivity in these metal phosphides and the nontrivial topological states, which requires further studies to establish.

In 1980, $Nb_2P_5$ was first synthesized by the high-pressure technique [1]. The electrical and magnetic properties were only measured down to ~77 K, which exhibits band paramagnetism and a metallic conduction behavior. Very recently, Liu et al. [2] had successfully grown high-quality $Nb_2P_5$single crystals and found a superconducting transition at around 2.6 K, by optimizing the high-pressure synthesis conditions.

A few of the physical properties, such as structural, high pressure crystal growth, electronic band structure, and superconducting state of $Nb_2P_5$, have been studied both theoretically and experimentally so far [1, 2]. Quite surprisingly, most of physical properties, e.g., elastic, electronic (DOS, charge density distribution, electron density difference), thermo-physical,

Mulliken population analysis, and optical properties of this interesting metallic phosphide $Nb_2P_5$ have not been explored at all. For example, analysis of the Cauchy pressure, tetragonal shear modulus, linear compressibility, Kleinman parameter, machinability index, hardness, acoustic velocity (both isotropic and anisotropic), acoustic impedance, anisotropy in elastic moduli and many more are still lacking. A number of thermo-physical properties, such as Debye temperature, melting temperature, interatomic bonding character, thermal expansion, heat capacity and minimum thermal conductivity (both isotropic and anisotropic) have not been investigated thoroughly yet. Besides energy dependence of the optical constants are still unknown. A thorough understanding of the elastic, mechanical, acoustic, electronic, thermal, bonding and optical constants spectra of $Nb_2P_5$ is crucial to unravel the potential of this compound for possible applications. We wish to bridge this research gap in this study and it constitutes the primary motivation of the present investigations.

The remaining part of this manuscript has been structured as follows: computational methodology is briefly described in Section 2. The computational results and their analyses are presented in Section 3. Finally, we have summarized the important features of this study and drawn pertinent conclusions in Section 4.

## 2. Computational method

All the structural, elastic, electronic, bonding, and optical calculations of $Nb_2P_5$were carried out by using plane wave pseudopotential approach based on the density functional theory (DFT) [16, 17] implemented in the quantum mechanical CASTEP (CAmbridge Serial Total Energy Package) simulation code [18]. The electronic exchange-correlation energy terms in the total energy calculations were incorporated via the generalized gradient approximation (GGA) as contained in the Perdew-Burke-Ernzerhof (PBE) scheme [19]. The coulomb potential energy arising due to the interaction between the valence electrons and ion cores were represented by Vanderbilt-type ultra-soft pseudopotential [20]. Use of ultra-soft pseudopotential saves substantial computational time with little compromise of computational accuracy. For Nb atoms $4s^24p^64d^45s^1$ electronic states and for B atoms $3s^23p^3$ electronic states were explicitly treated as valence electrons to perform the pseudo atomic calculations. The geometry optimization of $Nb_2P_5$, was performed using the Broyden–Fletcher–Goldfarb–Shanno (BFGS) minimization technique [21]. The cut off energy for expansion of electron wave function in the plane-wave basis set was taken as500 eV, which resulted in high level of convergence of the total energy calculations. The reciprocal-space integration over the Brillouin zone (BZ) was carried out using the Monkhorst-Pack scheme [22] with a mesh size of 6 x 8 x 6 k-points for $Nb_2P_5$. Geometry optimization of $Nb_2P_5$ was performed using total energy convergence tolerance of $10^{-5}$ eV/atom, maximum lattice point displacement within $10^{-3}$Å, maximum ionic Hellmann-Feynman force within 0.03 eVÅ$^{-1}$ and maximum stress tolerance of 0.05 GPa, with finite basis set corrections [23]. Gaussian smearing of 0.1 eV has been used. These selected tolerance levels have produced reliable estimates of structural, elastic and electronic band structure properties with an optimum computational time. All the first principles calculations have been performed at ground state with default temperature and pressure of 0 GPa and 0 K, respectively.

The stress-strain method is used in the calculations of single crystal elastic constants $C_{ij}$, of orthorhombic structure [24]. From symmetry considerations, an orthorhombic crystal has nine independent elastic constants ($C_{11}$, $C_{12}$, $C_{13}$, $C_{23}$, $C_{22}$, $C_{33}$, $C_{44}$, $C_{55}$, and $C_{66}$). The calculated values of single crystal elastic constants $C_{ij}$ allowed us to evaluate all the others elastic parameters, such as the bulk modulus (*B*) and shear modulus (*G*) using the Voigte-Reusse-Hill (VRH) approximation [25, 26].

The macroscopic electronic response of a material to incident electromagnetic wave can be fully described by the complex dielectric function $\varepsilon(\omega) = \varepsilon_1(\omega) + i\varepsilon_2(\omega)$. The imaginary part of the dielectric function, $\varepsilon_2(\omega)$, is obtained from the momentum matrix elements between the occupied and the unoccupied electronic states and has been calculated by directly using the CASTEP supported formula,

$$\varepsilon_2(\omega) = \frac{2e^2\pi}{\Omega\varepsilon_0}\sum_{k,v,c}|\langle\Psi_k^c|\hat{u}.\vec{r}|\Psi_k^v\rangle|^2\ \delta(E_k^c - E_k^v - E) \tag{1}$$

where, $\Omega$ is the unit cell volume, $\omega$ is the frequency of the incident light, e is the electronic charge, $u$ is the vector defining the polarization of the incident electric field, and $\Psi_k^c$ and $\Psi_k^v$ are the conduction and valence band wave functions at a given wave-vector k, respectively. This formula makes use of the calculated electronic band structure. The real part of the dielectric function, $\varepsilon_1(\omega)$, has been evaluated from the corresponding imaginary part of the dielectric function $\varepsilon_2(\omega)$ using the Kramers-Kronig transformation equation. Once the values of $\varepsilon_1(\omega)$ and $\varepsilon_2(\omega)$ are known, one can extract and analyze all the other optical constants such as the refractive index n(ω), absorption coefficient α(ω), energy loss-function L(ω), reflectivity R(ω), and optical conductivity σ(ω) from them [27].

To understand the bonding characteristics of solids, the Mulliken bond population analysis [28] can be employed. We have used a projection of the plane-wave states onto a linear combination of atomic orbital (LCAO) basis sets [29, 30] for $Nb_2P_5$. The Mulliken bond population analysis is carried out using the Mulliken density operator written on the atomic (or quasi-atomic) basis as follows:

$$P_{\mu\nu}^M(g) = \sum_{g'}\sum_{\nu'} P_{\mu\nu'}(g')S_{\nu'\nu}(g-g') = L^{-1}\sum_k e^{-ikg}\,(P_kS_k)_{\mu\nu'} \tag{2}$$

and the net charge on atom A can be defined as,

$$Q_A = Z_A - \sum_{\mu\in A} P_{\mu\mu}^M(0) \tag{3}$$

where $Z_A$ is the charge on the nucleus or atomic core (in simulations using the atomic pseudopotential).

## 3. Results and analysis

### *3.1 Structural properties*

The crystal structure of $Nb_2P_5$ is orthorhombic with space group Pnma (No. 62). The Wyckoff positions of Nb and P atoms in the unit cell are shown in Table 1. The schematic diagram of the crystal structure of $Nb_2P_5$ is shown in Figure 1. The cell contains eight Nb atoms and twenty *P* atoms. The results of first-principles calculations of ground state structural properties together with available experimental values [1, 2] are tabulated in Table 2. Our calculated values of lattice constants are in excellent agreement with the prior experimental results.

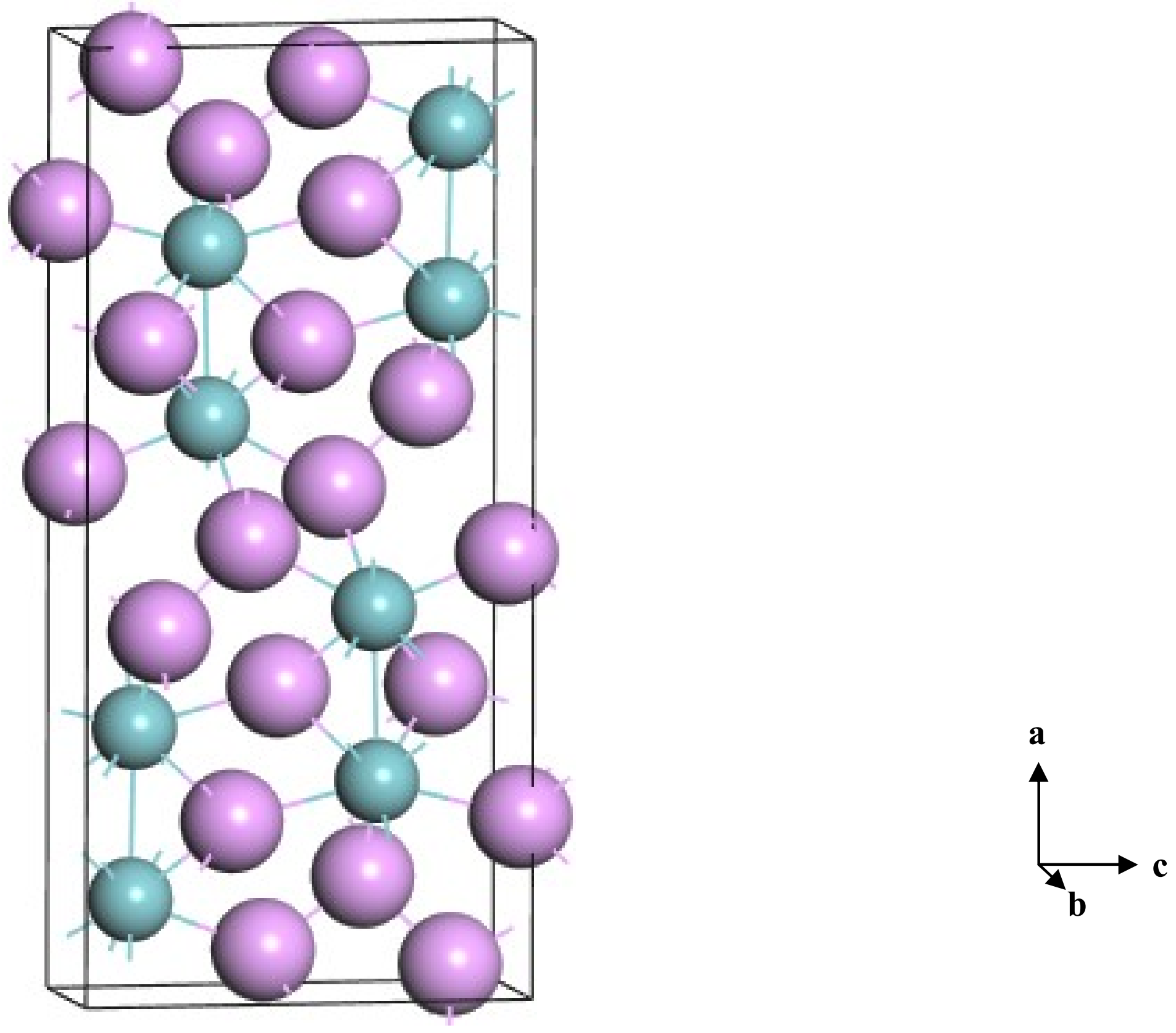

**Figure 1.** 3D schematic crystal structure of $Nb_2P_5$unit cell (The violet spheres denote the P atoms and the blue spheres denote the Nb atoms). The crystallographic directions are shown.

**Table 1**
Wyckoff positions of Nb and P atoms in a unit cell of $Nb_2P_5$.

| Species | *x* | *y* | *z* | Ref. |
|---|---|---|---|---|
| Nb(1) | 0.09506 | 0.25 | 0.16391 | |
| Nb(2) | 0.27343 | 0.25 | 0.17210 | |
| P(1) | 0.04016 | 0.25 | 0.46144 | |
| P(2) | 0.38426 | 0.75 | 0.18293 | [1]Exp. |
| P(3) | -0.02257 | 0.75 | 0.12131 | |
| P(4) | 0.31556 | 0.25 | 0.49344 | |
| P(5) | 0.18504 | 0.75 | 0.34794 | |

**Table 2**
Calculated and experimental lattice constants *a, b, c* (all in Å), equilibrium volume $V_o$ (Å$^3$),total number of atoms in the cell, total number of bonds *N*, and bulk modulus *B* (GPa) of $Nb_2P_5$.

| Compound | *a* | *b* | *c* | *b/a* | *c/a* | $V_o$ | No. of atoms | *N* | *B* | Ref. |
|---|---|---|---|---|---|---|---|---|---|---|
| | 16.78 | 3.33 | 7.90 | 0.10 | 0.47 | 441.90 | 28 | 60 | 142.66 | This |
| $Nb_2P_5$ | 16.74 | 3.35 | 7.91 | - | - | - | - | - | - | [1]Expt. |
| | 16.73 | 3.35 | 7.91 | - | - | - | - | - | - | [2] Expt. |

## *3.2 Mechanical and elastic properties*

Elastic properties of a crystalline compound correlate with many of its mechanical and physical qualities. Elastic constants are important to determine mechanical stability, that is, response of a material to the applied macroscopic stress. The calculated values of elastic constants of $Nb_2P_5$are listed in Table 3.For mechanical stability, according to Born-Huang conditions, a orthorhombic system requires to satisfy the following inequality criteria [31, 32]: $C_{11} > 0$, $C_{22} > 0$, $C_{33} > 0$, $C_{44} > 0$, $C_{55} > 0$, $C_{66} > 0$, $(C_{11}+C_{22}–2C_{12}) > 0$, $(C_{11}+C_{33}–2C_{13}) > 0$,$(C_{22}+C_{33}–2C_{23}) > 0$, $(C_{11}+C_{22}+C_{33}+2C_{12}+2C_{13}+2C_{23}) > 0$. All the elastic constants of $Nb_2P_5$ are positive and satisfy these mechanical stability criteria. This indicates that $Nb_2P_5$ is mechanically stable.

The elastic constants $C_{11}$ and $C_{33}$ represent the resistance to linear compression along [100] and [001] directions, respectively. Here it is seen that for $Nb_2P_5$ $C_{33}$ is larger than $C_{11}$, which indicates that the bonding strength/compressibility along [001] direction are stronger/lesser than those along [001] direction in $Nb_2P_5$. The uniaxial stiffness constants follow the sequence $C_{33} > C_{11} > C_{22}$, implying bonding along the b-direction is the weakest. The elastic constant $C_{44}$ represents the resistance to shear deformation with respect to a tangential stress applied to the (100) plane in the [010] direction of the compound. Here it is seen that for $Nb_2P_5$, $C_{44}$ is much lower than $C_{11}$ and $C_{33}$. This indicates that the compound is more easily deformed by a shear in comparison to a unidirectional stress along any of the three crystallographic directions. For the compound, $C_{44}$ is lower than $C_{66}$, which indicates that the shear along the (100) plane is easier relative to the shear along the (001) plane. Since $C_{11}+C_{12} > C_{33}$ for the compound, bonding in the (001) plane is more rigid elastically than that along the *c*-axis as well as the elastic tensile modulus is higher on the (001) plane than that along the *c*-axis. The tetragonal shear modulus, ($C' = \frac{C_{11}-C_{12}}{2}$) of a crystalline material is a measure of crystal's stiffness (the resistance to shear deformation by a shear stress applied in the (110) plane in the $[1\bar{1}0]$ direction). The tetragonal share modulus is also known as shear constant. The value of $C'$ is linked with the degree of stability of a crystal. A positive value of shear constant of a material indicates the stability with respect to tetragonal distortions; otherwise the system is expected to be dynamically unstable. The estimated value for $Nb_2P_5$ is given in Table 3. We see that the shear constant of $Nb_2P_5$ is 91.62 (positive) and the crystal is predicted to be dynamically stable.

The Kleinman parameter is another useful elastic parameter describing the relative position of the cation and anion sublattices under volume conserving distortions for which atomic positions are not fixed by crystal symmetry [33]. The measurement of the Kleinman parameter ($\zeta$), also known as internal strain parameter, is an indicator the stability of a compound against stretching and bending[34]. The Kleinman parameter ($\zeta$) is estimated from the following relation [35]:

$$\zeta = \frac{C_{11} + 8C_{12}}{7C_{11} + 2C_{12}} \tag{4}$$

The Kleinman parameter is a dimensionless parameter. The value of ζ generally lies between zero to one ($0 \leq \zeta \leq 1$). Here, the lower and upper limits of $\zeta$ ($\zeta = 0$ and $\zeta = 1$, respectively) are the indicator of the insignificant contribution of bond bending to resist the external stress and insignificant contribution of bond stretching/contracting to resist the external applied stress, respectively. The calculated value of $\zeta$ of $Nb_2P_5$ is 0.497. From which we predict that mechanical strength in $Nb_2P_5$ is mainly dominated by bond bending contribution over bond stretching or contracting.

The isotropic bulk modulus (*B*) and shear modulus (*G*) (by the Voigt-Reuss-Hill (VRH) method), Young's modulus (*Y*), Poisson's ratio (*v*) and hardness (*H*) of $Nb_2P_5$ is calculated using following well known equations [36-38]:

$$B_H = \frac{B_V + B_R}{2} \tag{5}$$

$$G_H = \frac{G_V + G_R}{2} \tag{6}$$

$$Y = \frac{9BG}{(3B + G)} \tag{7}$$

$$\nu = \frac{(3B - 2G)}{2(3B + G)} \tag{8}$$

$$H = \frac{(1 - 2\upsilon)Y}{6(1 + \upsilon)} \tag{9}$$

Isotropic shear modulus and bulk modulus are the gross measures of hardness of a material. The bulk modulus (*B*) is a measure of resistance to volume change by applied pressure, whereas the shear modulus (*G*) is a measure of resistance to reversible deformations due to shearing stress. For $Nb_2P_5$, larger value of *B* compared to *G* (Table 4) indicates that the mechanical strength will be limited by the shear deformation. The large value of shear modulus indicates pronounced directional bonding between atoms [39]. The covalent nature of a material increases with increasing value of Young modulus [40]. Young's modulus is a measure of the stiffness (resistance) of an elastic material to a change in its length [41, 42] and provides with a measure of thermal shock resistance. A material is stiff when the value of its Young's modulus is large. The calculated Young's modulus (Y) is comparatively large, therefore, $Nb_2P_5$is stiff material. The lattice thermal conductivity ($K_L$) and Young's modulus of a solid are related as: $K_L \sim \sqrt{Y}$ [43].

Pugh proposed that *G/B* ratio, which is known as Pugh's ratio that reflects the brittle/ductile nature of a solid [44-46]. The material is brittle, if the value of *G/B* is higher than 0.57; otherwise it would be ductile. In our case, the *G*/*B* value is 0.59 which indicates that the compound is expected show brittle behavior; though the brittleness is quite marginal. Besides, there is a typical relation between bulk and shear moduli for covalent materials which can be expressed roughly as $G \sim 1.1B$ [47]. For our case, the calculated values of bulk and shear moduli are 142.66 GPa and 84.84 GPa, respectively. This also indicates that the covalent bonding is dominant in$Nb_2P_5$.

The Poisson's ratio is a crucial parameter for studying number of properties of a compound, such as compressibility, brittle/ductile nature, and characteristic of bonding force. Poisson's ratio of a material can have the values between -1 (B << G) and ½ (B >> G) [38]. The Poisson's ratio, $\nu \sim 0.31$ is another useful indicator of brittle/ductile nature [48]. If $\nu \leq 0.33$, the material is expected to exhibit brittle fracture, otherwise it is ductile in nature. It is seen from Table 4 that the calculated value of *ν* of $Nb_2P_5$ stays below the critical value, suggesting that $Nb_2P_5$ is brittle in nature. For central-force solids, the Poisson's ratio is bounded by the lower and upper limits of 0.25 and 0.50, respectively [49, 50]. From the values of Poisson's ratio in Table 4, we can say that the interatomic forces of $Nb_2P_5$ are close to central in nature. The value of Poisson's ratio also indicates the presence of ionic and covalent bonding in a compound. For covalent and ionic materials the values of *ν* are typically ~0.10 and ~0.25, respectively [51]. The calculated Poisson's ratio of $Nb_2P_5$ is 0.25. This implies that ionic contribution is significant in $Nb_2P_5$. All together, it can be concluded that the binary phosphide $Nb_2P_5$ possesses mixed bonding character.

Another approach to determine brittle/ductile nature of a material is through the calculation of the Cauchy pressure $\left(C^{"} = C_{12} - C_{44}\right)$. If the Cauchy pressure is positive, the material is expected to be ductile; otherwise brittle [52]. The angular characteristics of atomic bonding in a material can be also described by Cauchy pressure [53]. Positive and negative values of the Cauchy pressure are linked with the presence of ionic bonding in a material and the presence of covalent bonding, respectively. It is seen that Cauchy pressure of $Nb_2P_5$ is positive suggesting that the compounds under consideration should be ductile in nature. The Pettifor's rule [53] states that, a material with large positive Cauchy pressures has more metallic bonds and thus become more ductile, on the other hand, if Cauchy pressure of the material is strongly negative, they possess more angular bonds, and thus exhibit more brittleness. Here, we should mention that positive Cauchy pressure found for $Nb_2P_5$ might be misleading because the corrections due to many body interactions among atoms and electron gas are not taken into account in determining the elastic constants [54]. Positive value of Cauchy pressure for $Nb_2P_5$ probably suggested that in addition to ionic and covalent bondings, these compounds have some metallic bondings as well. It is also worth noting that the Pugh's ratio of $Nb_2P_5$ lies very close to the boundary between brittleness and ductility.

Machinability of a material has become a useful topic in today's manufacturing industry. It is influenced by a number of variables, such as the inherent properties or characteristics of the work materials, cutting tool material, tool geometry, the nature of tool engagement with the work, cutting conditions, type of cutting, cutting fluid, and machine tool rigidity and its capacity [55]. The machinability index of materials defines cutting force, tool life and many more related characteristics. The machinability index, $\mu_M$ of a material can be calculated using following equation [56]:

$$\mu_M = \frac{B}{C_{44}} \qquad (10)$$

This particular parameter is also used for the measurement of plasticity [57-60] and dry lubricating property of a compound. Compounds with lower $C_{44}$ value give better dry lubricity. A compound with larger value of $B/C_{44}$ predicts excellent lubricating properties, lower feed forces, lower friction value, and higher plastic strain value. The $B/C_{44}$ values of $Nb_2P_5$ is 2.04, which indicates the presence of good level of machinability, comparable to many technologically important MAX phases and related compounds [61-65]. In order to understand elastic and plastic properties of a material, it is also important to know its hardness value. The calculated value of hardness [66] of $Nb_2P_5$ is 14.16 GPa. This implies that compound under study possesses reasonable hardness comparable to some of the binary intermetallic compounds [67].

**Table 3**

Calculated elastic constants, $C_{ij}$ (GPa), Cauchy pressure, ($C_{12}$-$C_{44}$) (GPa), tetragonal shear modulus, $C'$ (GPa), and Kleinman parameter (ζ) for $Nb_2P_5$ at $P$ = 0 GPa and $T$ = 0 K.

| Compound | $C_{11}$ | $C_{12}$ | $C_{13}$ | $C_{22}$ | $C_{23}$ | $C_{33}$ | $C_{44}$ | $C_{55}$ | $C_{66}$ | $C''$ | $C'$ | ζ | Ref. |
|---|---|---|---|---|---|---|---|---|---|---|---|---|---|
| $Nb_2P_5$ | 283.78 | 100.55 | 81.63 | 199.83 | 73.50 | 308.69 | 69.88 | 69.97 | 120.28 | 30.67 | 91.62 | 0.497 | This |

**Table 4**

The calculated isotropic bulk modulus $B$ (in GPa), shear modulus $G$ (in GPa), Young's modulus $Y$ (in GPa), Pugh's indicator $G/B$, Machinability index $B/C_{44}$, Poisson's ratio $v$ and hardness $H_V$ of $Nb_2P_5$ compounds deduced by Voigt, Reuss, and Hill (VRH) approximations.

| | $B$ | | | $G$ | | | $Y$ | $\frac{G_V}{G_R}$ | $\frac{B_V}{B_R}$ | $G/B$ | $\mu_M$ | ν | $H_V$ | Ref. |
|---|---|---|---|---|---|---|---|---|---|---|---|---|---|---|
| $Nb_2P_5$ | $B_V$ | $B_R$ | $B_H$ | $G_V$ | $G_R$ | $G_H$ | | | | | | | | |
| | 144.85 | 140.47 | 142.66 | 87.80 | 81.89 | 84.84 | 212.42 | 1.07 | 1.03 | 0.59 | 2.04 | 0.25 | 14.16 | This |

## *3.3 Elastic anisotropy*

Almost all the known crystalline materials are anisotropic. The directional dependency of mechanical properties of a crystalline solid can be explained by its elastic anisotropy. The study of anisotropic elastic properties of materials is essential because a number of important physical processes such as formation of micro-cracks in solids, motion of cracks, development of plastic deformations in crystals etc. are closely related to nature and extent of the anisotropy. For instance, the shear anisotropic factors measure degree of anisotropy in the bonding strength for atoms located in different planes. A proper explanation of these properties has significant implications in crystal physics as well as in applied engineering sciences. Therefore, it is crucial to calculate elastic anisotropy factors of $Nb_2P_5$ in details to understand their durability and possible applications under different external environments.

The shear anisotropic factors can be used to measure degree of anisotropy in the bonding between atoms in different planes. The shear anisotropy for an orthorhombic crystal can be quantified by three different factors [39, 68]:

The shear anisotropic factor for {100} shear planes between the ⟨011⟩ and ⟨010⟩ directions is,

$$A_1 = \frac{4C_{44}}{C_{11} + C_{33} - 2C_{13}} \quad (11)$$

The shear anisotropic factor for the {010} shear plane between ⟨101⟩ and ⟨001⟩ directions is,

$$A_2 = \frac{4C_{55}}{C_{22} + C_{33} - 2C_{23}} \tag{12}$$

and the shear anisotropic factor for the {001} shear planes between ⟨110⟩ and ⟨010⟩ directions is,

$$A_3 = \frac{4C_{66}}{C_{11} + C_{22} - 2C_{12}} \tag{13}$$

The calculated shear anisotropic factors of $Nb_2P_5$ are enlisted in Table 5. All three factors must be one ($A_1 = A_2 = A_3 = 1$), if the crystal is isotropic. Any other value (lesser or greater than unity) implies degree of anisotropy possessed by the crystal. The estimated values of $A_1$, $A_2$ and $A_3$ predict that the compound is moderately anisotropic. Our compound shows maximum anisotropy for $A_3$, which is 1.70.

The universal log-Euclidean index can be defined as [69, 70],

$$A^L = \sqrt{\left[ln\left(\frac{B^V}{B^R}\right)\right]^2 + 5\left[ln\left(\frac{C_{44}^V}{C_{44}^R}\right)\right]^2} \tag{14}$$

where, the Voigt and Reuss values of $C_{44}$ are obtained from [69]:

$$C_{44}^R = \frac{5}{3}\frac{C_{44}\,(C_{11} - C_{12})}{3(C_{11} - C_{12}) + 4C_{44}} \tag{15}$$

and

$$C_{44}^V = C_{44}^R + \frac{3}{5}\frac{(C_{11} - C_{12} - 2C_{44})^2}{3(C_{11} - C_{12}) + 4C_{44}} \tag{16}$$

The expression for $A^L$ is valid for any crystal symmetry. $A^L$ is closely related to the universal anisotropy index $A^U$. $A^U$ is a relative measure of anisotropy with respect to a limiting value. It does not represent an absolute level of anisotropy. For example, a compound with higher $A^U$ value does not necessarily indicate it has higher anisotropy, it just reveals the anisotropic nature. That is why we have calculated $A^L$ using the difference between the averaged stiffnesses $C^V$ and $C^R$, which is more appropriate. For a perfectly anisotropic crystal, $A^L$ is equal to zero. The value of $A^L$ for $Nb_2P_5$ is 2.97 indicating high level of anisotropy. The values of $A^L$ range between 0 to 10.26, and for almost 90% of the solids, $A^L < 1$. It has been argued that $A^L$ is also an indicator of the presence of layered/lamellar type of configuration [55, 69]. Materials with higher and lower $A^L$ values show strong layered and non layered structures, respectively. Therefore, the compound under study exhibits layered type of structural configuration, since the value of $A^L$ is comparatively higher.

The universal anisotropy index $A^U$, equivalent Zener anisotropy measure $A^{eq}$, anisotropy in compressibility $A^B$ and anisotropy in shear $A^G$ (or $A^C$) for the crystal with any symmetry are estimated using following standard equations [69, 71-73]:

$$A^U = 5\frac{G_V}{G_R} + \frac{B_V}{B_R} - 6 \geq 0 \tag{17}$$

$$A^{eq} = \left(1 + \frac{5}{12}A^U\right) + \sqrt{\left(1 + \frac{5}{12}A^U\right)^2 - 1} \tag{18}$$

$$A^B = \frac{B_V - B_R}{B_V + B_R} \tag{19}$$

$$A^G = \frac{G^V - G^R}{2G^H} \tag{20}$$

Because of its simplicity compared to the plurality of anisotropy factors defined for specific planes in crystals, universal anisotropy factor has become one of the most attractive anisotropy index. The concept of universal anisotropy index $A^U$ was introduced by Ranganathan and Ostoja-Starzewski [71], which provides a singular measure of anisotropy irrespective of the crystal symmetry. Unlike all other anisotropy indices, $A^U$ instituted the influence of the bulk to the anisotropy of a solid for the very first time. From Eqn. 14 it can be said that a larger fractional difference between the Voigt and Reuss estimated bulk or shear modulus would indicate a stronger degree of crystal anisotropy. From the values of $G_V/G_R$ and $B_V/B_R$ for $Nb_2P_5$, it is seen that $G_V/G_R$ has more influence on $A^U$ than that due to $B_V/B_R$. The universal anisotropy factor possesses only zero or positive value. The zero value of $A^U$ indicates that the crystal is isotropic and any deviation from this value suggests presence and level of anisotropy. $A^U$ for $Nb_2P_5$ is 0.39 which deviates from zero and the compound possesses anisotropy in elastic/mechanical properties.

For an isotropic crystalline material, $A^{eq} = 1$. The calculated value of $A^{eq}$ for $Nb_2P_5$ is 1.76 predicting that the compound is anisotropic. For an isotropic crystal, $A^G$ and $A^B$ are equal to zero. While any deviation from zero (positive) represents the degree of anisotropy. For $Nb_2P_5$, the value of $A^G$ is larger than $A^B$ (Table 5), which indicates that anisotropy in shear is larger than the anisotropy in compressibility.

The linear compressibility of an orthorhombic compound along $a$ and $c$ axis ($\beta_a$ and $\beta_c$) are evaluated from [74]:

$$\beta_a = \frac{C_{33} - C_{13}}{D} \quad \text{and} \quad \beta_c = \frac{C_{11} + C_{12} - 2C_{13}}{D} \tag{21}$$

with $D = (C_{11} + C_{12})C_{33} - 2(C_{13})^2$

From the calculated values (Table 5) one finds that compressibility along $a$ axis and $c$ axis indicates anisotropic nature, whereas anisotropic Bulk modulus clearly shows directional anisotropy in $Nb_2P_5$. Compared to all other anisotropy measurements, linear compressibility shows least anisotropy in $Nb_2P_5$.

**Table 5**

Shear anisotropic factors ($A_1$, $A_2$ and $A_3$), Zener anisotropy factor A, universal log-Euclidean index $A^L$, the universal anisotropy index $A^U$, equivalent Zener anisotropy measure $A^{eq}$, anisotropy in shear $A_G$ (or $A^C$) and anisotropy in compressibility $A_B$ (in GPa), linear compressibilities ($\beta_a$ and $\beta_c$) ($TPa^{-1}$) and their ratio ($\beta_c/\beta_a$) for $Nb_2P_5$at P = 0 GPa and T = 0 K.

| Compound | $A_1$ | $A_2$ | $A_3$ | $A^L$ | $A^U$ | $A^{eq}$ | $A_G$ | $A_B$ | $\beta_a$ | $\beta_c$ | $\frac{\beta_c}{\beta_a}$ | Layered | Ref. |
|---|---|---|---|---|---|---|---|---|---|---|---|---|---|
| $Nb_2P_5$ | 0.65 | 0.77 | 1.70 | 2.97 | 0.39 | 1.76 | 0.035 | 0.015 | 0.002 | 0.002 | 1 | Yes | This |

From pressure dependent lattice parameter measurements it is easy to obtain the bulk modulus of a solid along different crystallographic axes. It is however simpler to calculate the bulk modulus along the crystallographic axes from the single crystal elastic constants. The uniaxial bulk modulus along $a$, $b$ and $c$ axis and anisotropies of the bulk modulus can be defined as follows [68]:

$$B_a = a\frac{dP}{da} = \frac{\Lambda}{1+\alpha+\beta} \quad ; \quad B_b = a\frac{dP}{db} = \frac{B_a}{\alpha} \quad ; \quad B_c = c\frac{dP}{dc} = \frac{B_a}{\beta} \qquad (22)$$

and

$$A_{B_a} = \frac{B_a}{B_b} = \alpha \quad ; \quad A_{B_c} = \frac{B_c}{B_b} = \frac{\alpha}{\beta} \qquad (23)$$

where, $\Lambda = C_{11} + 2C_{12}\alpha + C_{22}\alpha^2 + 2C_{13}\beta + C_{33}\beta^2 + 2C_{33}\alpha\beta$

For orthorhombic crystals $\alpha$ and $\beta$ can be defined as,

$$\alpha = \frac{(C_{11} - C_{12})(C_{33} - C_{13}) - (C_{23} - C_{13})(C_{11} - C_{13})}{(C_{33} - C_{13})(C_{22} - C_{12}) - (C_{13} - C_{23})(C_{12} - C_{23})}$$

and

$$\beta = \frac{(C_{22} - C_{12})(C_{11} - C_{13}) - (C_{11} - C_{12})(C_{23} - C_{12})}{(C_{22} - C_{12})(C_{33} - C_{13}) - (C_{12} - C_{23})(C_{13} - C_{23})}$$

where, $A_{B_a}$and $A_{B_c}$are anisotropies of bulk modulus along the *a* axis and *c* axis with respect to *b* axis, respectively. The calculated results are listed in Table 6. For$Nb_2P_5$, $A_{B_a} = 1$ and $A_{B_b} \neq 1$, which indicates anisotropy in axial bulk modulus. Bulk modulus along *b*-axis is smaller than those along *a*- and *c*-axis. Also, the values of uniaxial Bulk modulus are different from the isotropic bulk modulus and are much larger. This originates from the fact that the pressure in a state of uniaxial strain for a given crystal density generally differs from the pressure in a state of hydrostatic stress at the same density of the crystal [39]. These results are also in good accord with the result we got for linear compressibility of $Nb_2P_5$.

**Table 6**
Anisotropies of bulk modulus along different axes of $Nb_2P_5$ compound.

| Compound | $B_a$ | $B_b$ | $B_c$ | $A_{B_a}$ | $A_{B_c}$ | Ref. |
|---|---|---|---|---|---|---|
| $Nb_2P_5$ | 821.93 | 424.11 | 733.21 | 1.94 | 1.73 | This |

### *3.4 Acoustic velocities and its anisotropy*

Sound velocity of a material is an important property which is connected with its thermal and electrical conductivity. Moreover, interest in studying acoustic behavior of a material has become a matter of notable concern in physics, materials science, seismology, geology, musical instrument designing, and medical sciences in recent days. A crystal structure with higher sound velocity shows higher thermal conductivity, $k = \frac{1}{3}C_v l v$. The velocity of transverse and longitudinal sound waves traversing through a crystalline solid can be related by the bulk and shear moduli using following equations [75]:

$$v_t = \sqrt{\frac{G}{\rho}} \quad \text{and} \quad v_l = \sqrt{\frac{B + 4G/3}{\rho}} \tag{24}$$

where $\rho$ is the mass-density of the solid.

The average sound velocity $v_a$of a polycrystalline system is evaluated from the transverse and longitudinal sound velocities using following equation [75]:

$$v_a = \left[\frac{1}{3}\left(\frac{2}{v_t^3} + \frac{1}{v_l^3}\right)\right]^{-\frac{1}{3}} \tag{25}$$

The estimated acoustic velocities have been summarized in Table 7.

The acoustic impedance of a material is an important factor of transducer design, noise reduction in aircraft engine, industrial factories and many underwater acoustic applications. When sound travels through one material and then encounters a different material, the amount of transmitted and reflected sound energy at the interface depends on the difference in acoustic impedances. Therefore, most of the sound gets transmitted, if the two impedances are about equal, but most of it is reflected back rather than being transmitted, if the impedance difference is high. The acoustic impedance of a material can be evaluated by the following expression [55, 76]:

$$Z = \sqrt{\rho G} \tag{26}$$

The unit of acoustic impedance is the Rayl: 1 Rayl = kgm$^{-2}$s$^{-1}$ = 1 Nsm$^{3}$. The above equation indicates that materials with higher density and higher shear modulus have higher acoustic impedance.

The intensity of sound radiation is another important design parameter for the construction of sound boards, such as the front plate of a violin, the sound board of a harpsichord and the panel of a loudspeaker. The radiated intensity, $I$, is proportional to the surface velocity for a given driving function, this scales with modulus of rigidity and density as [55, 76]:

$$I \approx \sqrt{G/\rho^3} \tag{27}$$

where, $\sqrt{G/\rho^3}$ is also known as the radiation factor. A high value of radiation factor is used by instrument makers to select materials for suitably designed sound boards. For example, spruce (8.6 m$^4$/kg.s), widely used for the front plates of violins, has a particularly high value—nearly twice that of maple (5.4 m$^4$/kg.s), which is used for the back plate, the function of which is to reflect, not radiate. The evaluated radiation factor for $Nb_2P_5$isgivenin Table 7.

**Table 7**

Density $\rho$ (g/cm$^3$), transverse sound velocity $v_t$(ms$^{-1}$), longitudinal sound velocity $v_l$ (ms$^{-1}$), average sound wave velocity $v_a$ (ms$^{-1}$), acoustic impedance $Z$ (Rayl) and radiation factor $\sqrt{G/\rho^3}$ (m$^4$/kg.s) of $Nb_2P_5$.

| Compound | $\rho$ | $v_t$ | $v_l$ | $v_a$ | $Z$ ($\times 10^6$) | $\sqrt{G/\rho^3}$ | Ref. |
|---|---|---|---|---|---|---|---|
| $Nb_2P_5$ | 5.12 | 4070.34 | 7060.79 | 4482.96 | 20.84 | 0.795 | This |
| | 5.08 | - | - | - | - | - | [1]$^{\text{Expt.}}$ |

The velocity of propagation of sound (longitudinal and transverse) waves in a material does not depend on its frequency and the dimension of the material but only on its nature. Each atom in a system has three modes of vibrations, one longitudinal and two transverse modes. The anisotropic properties of sound velocities suggest elastic anisotropy in a crystal and vice versa. In the case of anisotropic crystal, the pure longitudinal and transverse modes are found only along certain crystallographic directions. In all other direction the sound propagating modes are the quasi-transverse or quasi longitudinal waves. The pure transverse and longitudinal modes can be found for [100], [010], and [001] directions in an orthorhombic crystal. For orthorhombic crystal, the acoustic velocities along these principle directions can be expressed as [77]:

*[100]:*

$$[100]v_l = \sqrt{C_{11}/\rho}; [010]v_{t1} = \sqrt{C_{66}/\rho}; [001]v_{t2} = \sqrt{C_{55}/\rho}$$

*[010]:*

$$[010]v_l = \sqrt{C_{22}/\rho}\ ; [100]v_{t1} = \sqrt{C_{66}/\rho}; [001]v_{t2} = \sqrt{C_{44}/\rho} \quad (28)$$

*[001]:*

$$[001]v_l = \sqrt{(C_{33})/\rho}; [100]v_{t1} = \sqrt{C_{55}/\rho}; [001]v_{t2} = \sqrt{C_{44}/\rho}$$

where $v_{t1}$ and $v_{t2}$ are the first transverse mode and the second transverse mode, respectively.The calculated acoustic velocities in the directions for $Nb_2P_5$ are shown in Table 8.

It is obvious that both longitudinal and transverse sound velocities are correlated to elastic constants and crystal density, a compound with small density and large elastic constants has large sound velocities. The elastic anisotropy in a crystal indicates anisotropic properties in sound velocities. Here, $C_{11}$, $C_{22}$, and $C_{33}$ determine the longitudinal sound velocities along [100], [010] and [001] directions, respectively, and $C_{44}$, $C_{55}$, and $C_{66}$ correspond to the transverse mode.

**Table 8**
Anisotropic sound velocities (in $ms^{-1}$) of $Nb_2P_5$ along different crystallographic directions.

| Propagation directions | | $Nb_2P_5$ |
|---|---|---|
| [100] | $[100]v_l$ | 7444.16 |
| | $[010]v_{t1}$ | 4846.38 |
| | $[001]v_{t2}$ | 3696.37 |
| [010] | $[010]v_l$ | 6246.66 |
| | $[100]v_{t1}$ | 4846.38 |
| | $[001]v_{t2}$ | 3693.89 |
| [001] | $[001]v_l$ | 7763.99 |
| | $[100]v_{t1}$ | 3696.37 |
| | $[001]v_{t2}$ | 3693.89 |

## 3.5 Thermal properties

### 3.5.1 Debye temperature

Debye temperature ($\Theta_D$) is one of the most fundamental thermo-physical parameters of solids, closely related to large number of physical properties namely, thermal conductivity, lattice vibration, interatomic bonding, melting temperature, coefficient of thermal expansion, phonon specific heat and elastic constants. For a solid it also distinguishes between high- and low-temperature region in the phonon dynamics. Generally, a compound with stronger interatomic bonding strength, lower average atomic mass, higher melting temperature, greater hardness and higher mechanical wave velocity has larger Debye temperature. If $T > \Theta_D$, all modes are expected to have energy $k_B\Theta_D$. Whereas if $T < \Theta_D$, high-frequency modes are frozen [78]. At low temperatures the Debye temperature estimated from elastic constants is the same as that determined from specific heat measurements, since at low temperature the vibrational excitation arises solely from acoustic modes of lattice vibration). Under this condition the Debye temperature is proportional to the average sound velocity and can be obtained from the following equation [75, 79]:

$$\Theta_D = \frac{h}{k_B}\left(\frac{3n}{4\pi V_0}\right)^{1/3} \nu_a \tag{29}$$

where, $h$ is Planck's constant, $k_B$ is the Boltzmann's constant, $V_0$ is the volume of unit cell and $n$ is the number of atoms within the unit cell. This methodology developed by Anderson [75, 79] yields reliable estimates of $\Theta_D$ for solids belonging to different types and crystal symmetries [80-84]. The calculated value of the Debye temperature of $Nb_2P_5$ is listed in Table 9. Debye temperature of $Nb_2P_5$ is obtained as 328.25 K.

### 3.5.2 Melting temperature

The melting temperature is another interesting and important parameter for solids. It sets the limit in the high temperature applications. The bonding energy and thermal expansion of a crystalline material correlate with the melting temperature. High value of melting temperature indicates strong atomic bonding, high value of bonding energy and low value of thermal expansion. The melting temperature of a material also gives indication about the temperature up to which the material can be used continuously without oxidation, chemical change, and excessive distortion. The elastic constants can be used to calculate the melting temperature $T_m$ of solids using following equation [85]:

$$T_m = 354K + (4.5K/GPa)\left(\frac{2C_{11} + C_{33}}{3}\right) \pm 300K \tag{30}$$

The calculated melting temperature of $Nb_2P_5$ is listed in Table 9. The melting temperature of $Nb_2P_5$ is 1668.39 ± 300 K which indicates that it is a good candidate material for high temperature applications. Melting temperature of $Nb_2P_5$ is comparable to some of the technologically important MAX phase nanolaminates [86] suitable for use at high temperatures.

### 3.5.3 Thermal expansion coefficient and heat capacity

The thermal expansion coefficient (TEC) of a material is connected with number of other physical properties, like, specific heat, thermal conductivity, temperature variation of the energy band gap and electron effective mass. Thermal expansion coefficient ($\alpha$) is an intrinsic thermal property of a material and it is important for epitaxial growth of crystals. The thermal expansion coefficient of a material can be estimated using the following equation [55, 76]:

$$\alpha = \frac{1.6 \times 10^{-3}}{G} \tag{31}$$

where, $G$ is the shear modulus (in GPa). Thermal expansion coefficient is inversely related to melting temperature: $\alpha \approx 0.02/T_m$[76, 87]. Computed thermal expansion coefficient of $Nb_2P_5$ is disclosed in Table 9.

Heat capacity is another intrinsic thermodynamic parameter of a compound. In general, a compound with higher heat capacity has higher thermal conductivity and lower thermal diffusivity. The heat capacity per unit volume is the change in thermal energy per unit volume in a material per Kelvin change in temperature. The heat capacity per unit volume was determined by using following equation [55, 76]:

$$\rho C_P = \frac{3k_B}{\Omega} \tag{32}$$

where, $N = 1/\Omega$ is the number of atoms per unit volume. The heat capacity per unit volume of $Nb_2P_5$ is listed in Table 9.

Phonons play an important role in determining variety of physical properties of condensed matter, such as heat capacity, thermal conductivity and electrical conductivity. The wavelength of the dominant phonon is the wavelength, $\lambda_{dom}$, at which the phonon distribution attains a peak. The wavelength of the dominant phonon for $Nb_2P_5$ at 300K has been calculated by the following relationship [87]:

$$\lambda_{dom} = \frac{12.566 v_a}{T} \times 10^{-12} \tag{33}$$

where, $v_a$ is the average sound velocity in $ms^{-1}$, T is the temperature in K. A material with higher average sound velocity, higher shear modulus, lower density has higher wavelength of the dominant phonon. The estimated value of $\lambda_{dom}$ in meter is listed in Table 9.

**Table 9**
The Debye temperature $\Theta_D$ ($K$), thermal expansion coefficient α ($K^{-1}$), wavelength of the dominant phonon mode at 300K $\lambda_{dom}$ (m), melting temperature $T_m$ ($K$) and heat capacity per unit volume $\rho C_P$ ($J/m^3.K$) of $Nb_2P_5$.

| Compound | $\Theta_D$ | α ($10^{-5}$) | $\lambda_{dom}$ $(10^{-12})$ | $T_m$ | $\rho C_P$ $(10^6)$ |
|---|---|---|---|---|---|
| $Nb_2P_5$ | 328.25 | 1.886 | 187.78 | 1668.39 | 2.63 |

### *3.5.4 Minimum thermal conductivity and its anisotropy*

At temperatures above the Debye temperature ($T > \Theta_D$), thermal conductivity of a material approaches to its minimum value known as minimum thermal conductivity ($k_{min}$). The study of minimum thermal conductivity of materials has become desirable in identifying and development of compounds to be used for thermal barrier coatings (TBCs). One of the conspicuous features of the minimum thermal conductivity is that it is independent of the presence of defects inside the crystal, such as dislocations, individual vacancies and long-range strain fields associated with inclusions and dislocations. This is mostly because these defects affect phonon transport over length scales much more than the interatomic spacing and at high temperatures the phonon mean free path becomes significantly smaller than this length scale. Using Debye model, Clarke deduced the following formula for the calculation of minimum thermal conductivity $k_{\min}$ of a compound at high temperatures [87]:

$$k_{min} = k_B v_a (V_{atomic})^{-2/3} \tag{34}$$

In this equation, $k_B$ is the Boltzmann constant, $v_a$ is the average sound velocity and $V_{atomic}$ is the cell volume per atom [87]. A compound with higher acoustic velocity, minimum phonon mean free path and Debye temperature has higher minimum thermal conductivity.

The calculated value of isotropic minimum thermal conductivity for $Nb_2P_5$ using obtained Debye temperature is enlisted in Table 10.

There are three different modes of transmission of heat through solids. Firstly, by thermal vibrations of atoms, secondly, by the movement of free electrons in metals, and finally, if the solid is transparent, by radiation. The propagation of elastic waves dominates transmission of heat through thermal vibrations. An elastically anisotropic material also possesses anisotropic minimum thermal conductivity. The acoustic velocities of a compound along different crystallographic directions determine its anisotropy in minimum thermal conductivity. We have calculated the minimum thermal conductivities of $Nb_2P_5$ along different crystallographic directions using the method proposed by Cahill [88]:

$$k_{min} = \frac{k_B}{2.48} n^{2/3}(v_l + v_{t1} + v_{t2}) \tag{35}$$

and $$n = N/V$$

where, $k_B$ is the Boltzmann constant, $n$ is the number of atoms per unit volume and $N$ is total number of atoms in the cell having a volume $V$.

The minimum thermal conductivity of $Nb_2P_5$ in the [100], [010] and [001] directions using anisotropic acoustic velocities are summarized in Table 10. The results reveal that the minimum thermal conductivity of $Nb_2P_5$ has significant anisotropy along the three directions. Minimum thermal conductivity along [010] has the lowest value in comparison with the two other directions.

To compare these two methods, we have also calculated the isotropic minimum thermal conductivity of $Nb_2P_5$ using Cahill's method. The estimated results are disclosed in Table 10. The isotropic minimum thermal conductivity using Cahill's and Clarke's methods are 1.38 and 1.01, respectively. Clearly, Cahill's model predicts higher $K_{min}$ than that from the Clarke's model.

**Table 10**
The number of atom per mole of the compound $n$ ($m^{-3}$), minimum thermal conductivity (W/m.K) of $Nb_2P_5$ along different directions evaluated by Cahill's and Clarke's method.

| Compound | $n$ ($10^{28}$) | $[100]k_{calc.}^{Min}$ | $[010]k_{calc.}^{Min}$ | $[001]k_{calc.}^{Min}$ | $k_{min}$ | | Ref. |
|---|---|---|---|---|---|---|---|
| | | | | | Cahill | Clarke | |
| $Nb_2P_5$ | 6.35 | 1.45 | 1.34 | 1.37 | 1.38 | 1.01 | This |

## *3.6 Electronic properties*

### *3.6.1 Band structure*

The electronic band structure is one of the most important aspects in solid state physics that can explain various materials properties electrical conductivity, electronic thermal conductivity, electronic heat capacity, Hall effect, magnetic properties and optoelectronic characteristics. The calculated electronic band structure, as a function of energy ($E$-$E_F$), along different high symmetry directions ($\Gamma$-$X$-$S$-$Y$-$\Gamma$-$S$) in the first BZ at zero pressure and temperature is shown in Fig. 2. The horizontal broken line marks the Fermi level ($E_F$). The total number of bands for $Nb_2P_5$ is 215.

From Fig. 2 it is noticed that there is no band gap at the Fermi level, due to the overlap of valence band and conduction band. This indicates that $Nb_2P_5$ is metallic in nature. The electronic dispersion curves show that the energy bands around the Fermi level are mainly derived from the 4*d* electronic states of the Nb atom. Some contributions of the P 3*d* orbitals are also noted. This indicates that the Nb4*d* electrons should dominate the charge transport properties of $Nb_2P_5$.

**(a)**

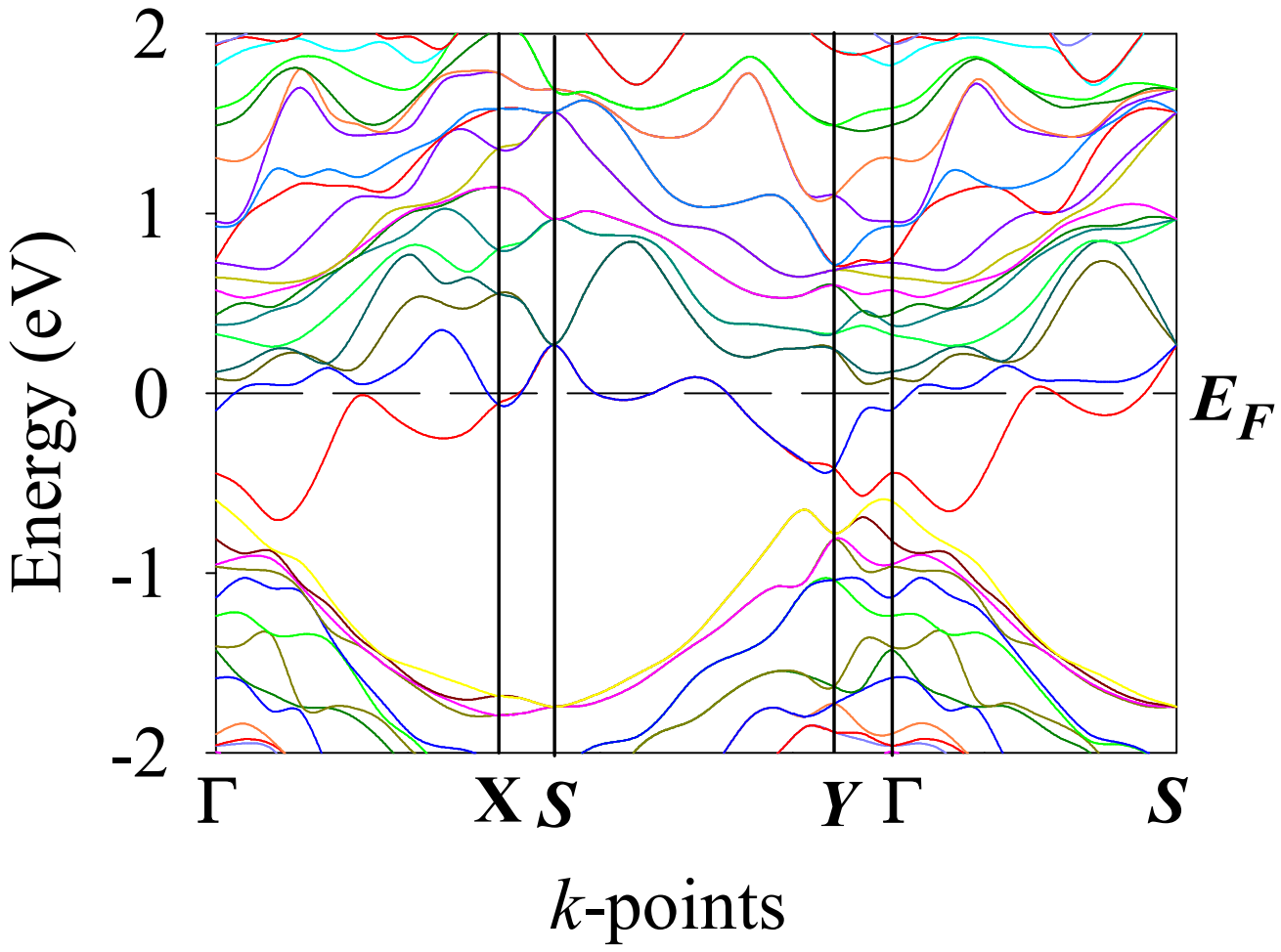

**(b)**

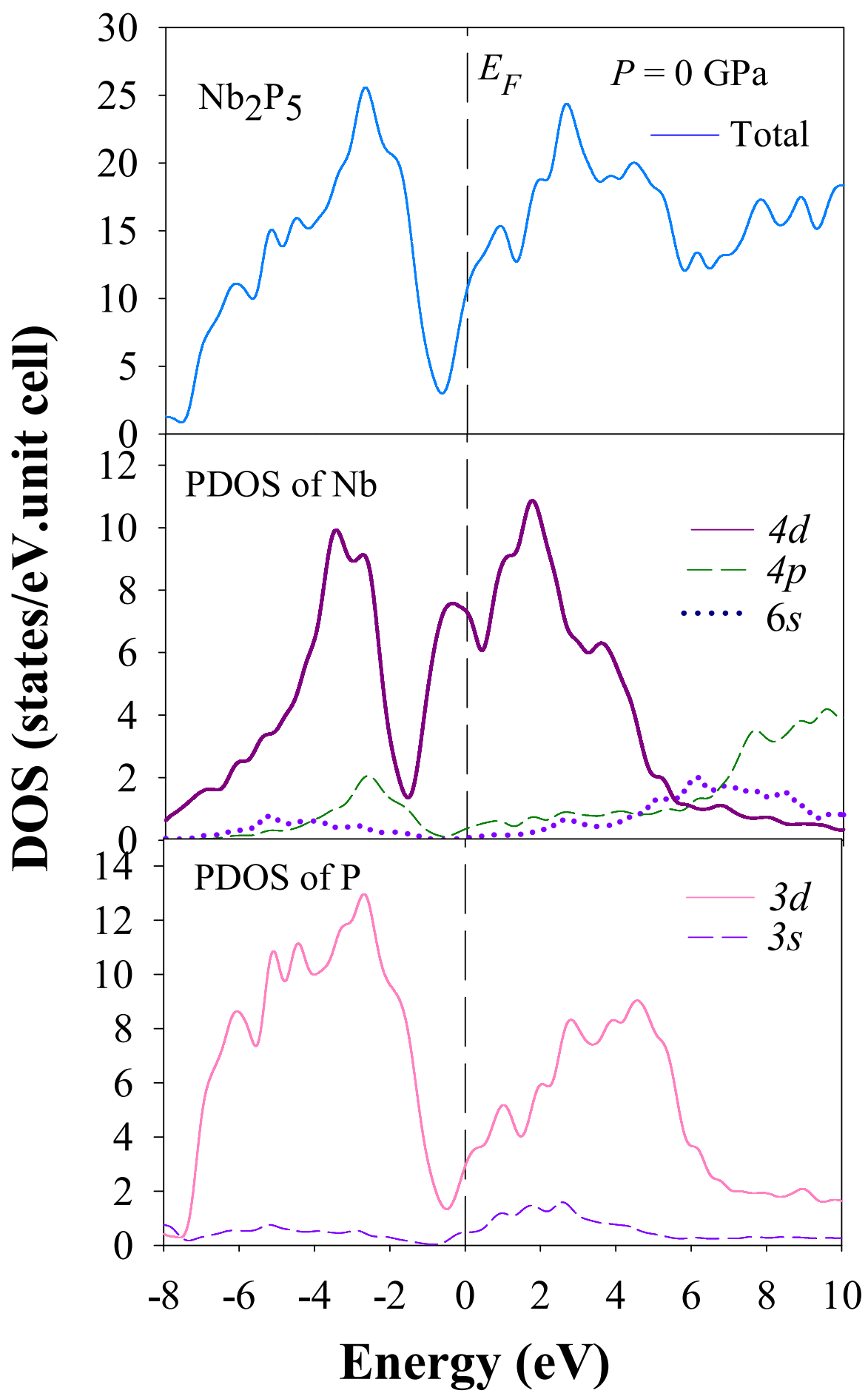

**Figure 2.** Electronic band structure of $Nb_2P_5$ along several high symmetry directions of the Brillouin zone at P = 0 GP.

**Figure 3.** Total and partial electronic density of states (TDOS and PDOS, respectively) of $Nb_2P_5$ as a function of energy. The Fermi level is placed at zero energy.

### *3.6.2 Density of states (DOS)*

In solid state physics, the electronic energy density of states or simply DOS of a system describes the number of electronic states available to be occupied at each energy level per unit energy interval. Almost all of the electronic and optical properties of crystalline solids are linked with the structure of the DOS. Moreover the DOS of a material is important to understand the contribution of each atom to bonding and antibonding states. The calculated total and partial density of states (TDOS and PDOS, respectively) of $Nb_2P_5$ at zero pressure and temperature is shown in Figure 3. The vertical broken line denotes the Fermi level, $E_F$. The non-zero value of TDOS at the Fermi level indicates that the compound under study will exhibit metallic electrical conductivity. We have also calculated the PDOS of Nb and P atoms to understand their contribution to the TDOS and chemical bonding. The dominant contribution to the TDOS in the vicinity of $E_F$ comes from Nb 4*d* and P 3*d* states. At the Fermi level their values are 7.35 and 3.07 states per eV per unit cell. TDOS value of $Nb_2P_5$ at the Fermi level is~10.82 states per eV per unit cell. The main contribution comes from Nb 4*d* state which is about 70% of the TDOS. The chemical and mechanical stability of $Nb_2P_5$ are also mainly influenced by the properties of Nb 4*d* electronic states. There are large peaks in the TDOS close to the $E_F$ located at -2.66 eV and 2.63 eV. These peak positions and the position of the Fermi level are quite sensitive to the lattice distortion, doping, stress etc. Therefore, there is significant scope of band engineering in $Nb_2P_5$. Close to the Fermi level, there is significant hybridization between these electronic states. Such hybridization near the Fermi energy is often indicative of formation of covalent bonding. From TDOS curves, we can see that there is a large bonding peak below and a smaller antibonding peak above the Fermi level. These peaks are located at -2.66 eV and 2.55 eV and are formed by the hybridization between the Nb 4*d* orbitals and P 3*d* orbitals. Thus one expects covalent bondings between these two atomic species.

The TDOS can be decomposed into three energy regions: the lowest energy region is due to P 3*s* electrons; the region from -5 to 6 eV are stemming mainly from Nb 4*d* and P 3*d* electrons; the region above this is mainly from the Nb 4*p* electrons.

Electronic and structural stability of a compound is related to the position of Fermi level with respect to the peaks in the TDOS nearby [89, 90].The presence of a deep valley in the TDOS curve in the vicinity of Fermi level is known as pseudogap or quasi-gap. This gap in the TDOS separates bonding states from nonbonding/antibonding electronic states. It is related to the electronic stability of a solid [91, 92]. For $Nb_2P_5$, Fermi level falls in the antibonding region (see Fig. 3).The origin of the pseudogap or the quasi-gap is usually responsible for two mechanisms [91, 93]. One is due to charge transfer (ionic origin) and the other one is due to hybridization among atomic orbitals. Moreover, presence of a pseudogap around the Fermi level often indicates towards the presence of the directional bonding [94] which facilates the formation of covalent bonding and enhances the mechanical strength of materials. To be specific, around the Fermi level, the strong bonding hybridization is originated mainly from the Nb 4*d* states with P 3*d* states, which thus forms the directional covalent bonding between Nb and P. This agrees with the electronic charge density (distribution and difference) mapping and Mulliken bond population analysis results (presented in Section 3.6.3 and 3.7, respectively).

Apart from structural stability, density of states of a compound provides with wealth of other information. (1) Firstly if the Fermi level lies exactly at the pseudogap, it indicates that the compound will have larger ordering energy and high melting point. (2) If the Fermi level lies to the right of the pseudogap, i.e., in the antibonding region, which refers to an unstable state and the system will show the tendency towards electronic instability and be in the disordered or glassy state [95]. Furthermore, if $E_F$ is close to the antibonding peak, $N(E_F)$ has high value, the system can exhibit fairly high superconducting transition temperature, high Pauli paramagnetic susceptibility and electronic specific heat coefficient [96]. (3) If the Fermi level lies in the high energy of antibonding region the compound will be in the unstable state and in order to minimize the energy, the system will try to go to some other lower energy configuration. (4) On the other hand, if the Fermi level lies to the left of the pseudogap, i.e., in the bonding states, not all bonding states are completely filled and additional electrons are needed to make the structure more stable [96]. For the stability of a compound, the charge interaction among bonding atoms is crucial, the compound possesses higher number of bonding electrons should be structurally more stable [97, 98].

The electron-electron interaction parameter, also known as the Coulomb pseudopotential parameter, of a compound can be estimated using the following relation [99]:

$$\mu^* = \frac{0.26N(E_F)}{1 + N(E_F)} \tag{36}$$

The total density of states at the Fermi level for $Nb_2P_5$ is 11.04 states/eV.unit cell. The electron-electron interaction parameter of $Nb_2P_5$ is found to be 0.24. This repulsive Coulomb pseudopotential reduces the superconducting critical temperature, $T_c$, of superconducting compounds [95, 99, 100].

### *3.6.3 Electronic charge density distribution*

To gain further understanding of the bonding nature of $Nb_2P_5$, the charge distribution around the atoms within the crystal has been investigated. The electronic charge density (e/Å$^3$) distribution of $Nb_2P_5$ had been calculated in different crystal planes. Figure 4 shows the electronic charge density distribution of $Nb_2P_5$ in the (100) and (111) planes. The color scales on the right hand side of charge density maps illustrate the total electron density (blue color indicates high charge (electron) density and red color indicates low charge (electron) density). The charge density distribution map shows weak signatures of covalent bonding between Nb-Nb, Nb-P and P-P atoms. The Mulliken bond population analysis (Section 3.7) also agrees with these findings. For (100) plane, the charge density difference between Nb and P atoms is negligible. On the other hand, the charge density distribution in (111) plane shows different result (Fig. 4b). It can be seen that electron density are centered more around the four Nb atoms compared to other and small fraction of charge is distributed along Nb-Nb bond and Nb-P bond, indicating dominant ionic and weak covalent character of bonds. However the bonding nature in (111) plane of $Nb_2P_5$ cannot be described as an exclusively ionic type. These plots show the simultaneous existence of ionic and covalent bonding between Nb-Nb, Nb-P and P-P atoms. P atoms are electron deficient in both planes and almost every atom (Nb and P) has nearly spherical charge distribution. The overall features are consistent with the DOS curves (Fig. 3a), which shows strong hybridization between Nb-4*d* and P-3*d* in $Nb_2P_5$. Uniform low density charge distribution far from the atomic sites in Figs. 4a and 4b imply that some contribution due to metallic bonding also exists.

**(a)** **(b)**

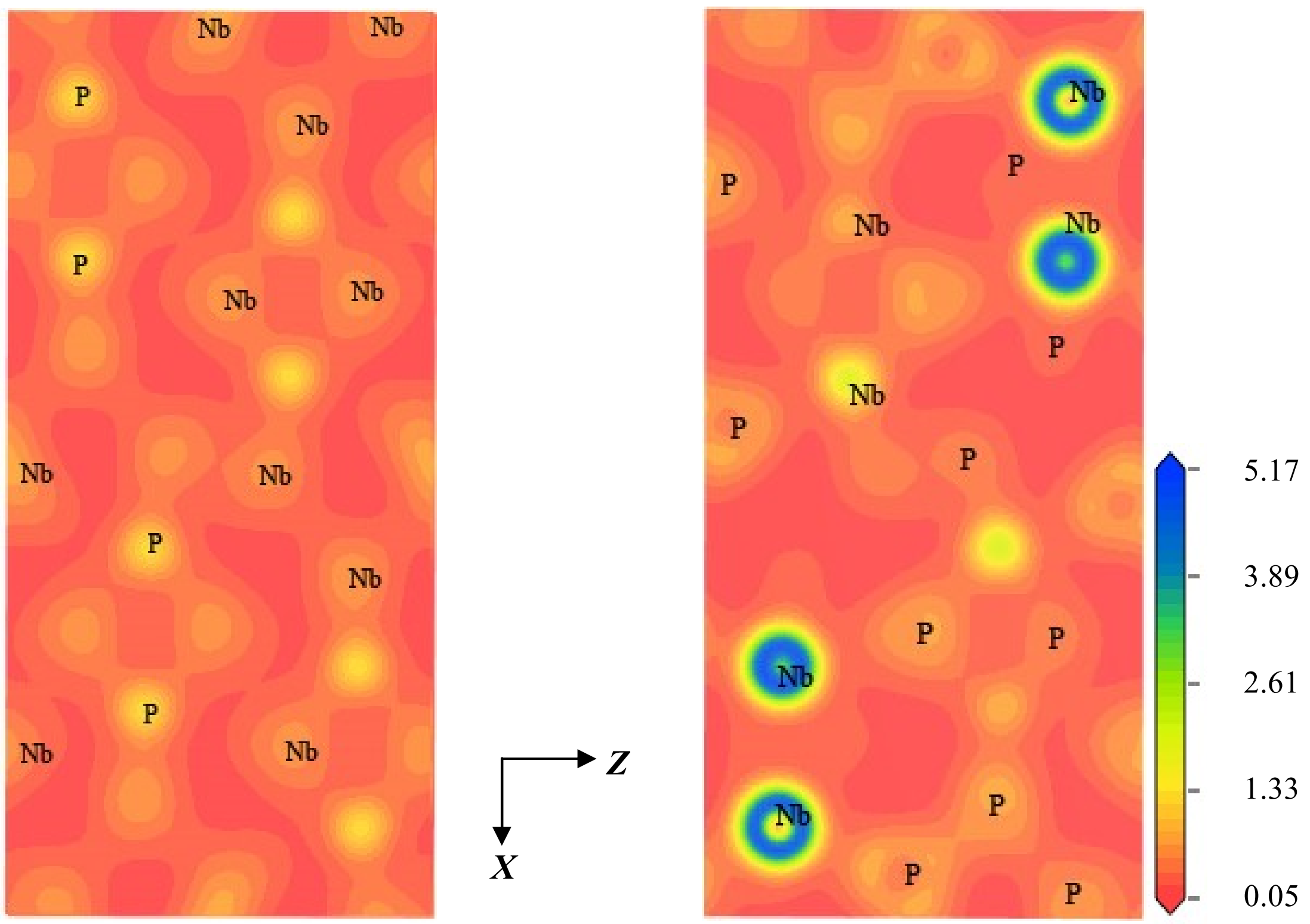

**Figure 4.** The electronic charge density distribution map of$Nb_2P_5$ (a) in the (100) plane and (b) in the (111) plane.

### *3.6.4 Electron density difference*

We have also calculated electron density difference of $Nb_2P_5$ in (001) and (111) planes. Fig. 5 shows the electron density difference of $Nb_2P_5$ in (111) plane. Electron density difference is another useful way to analyze bonding characters of compounds. The color scales on the right hand side illustrate the electron density difference. The scale of this map is from -0.394 to 0.236 electrons/Å$^3$. The extreme negative or positive electron density relative to the atom electron density is represented by red or blue regions, respectively. The main features of charge density distribution and electron density difference are in complete agreement with each other.

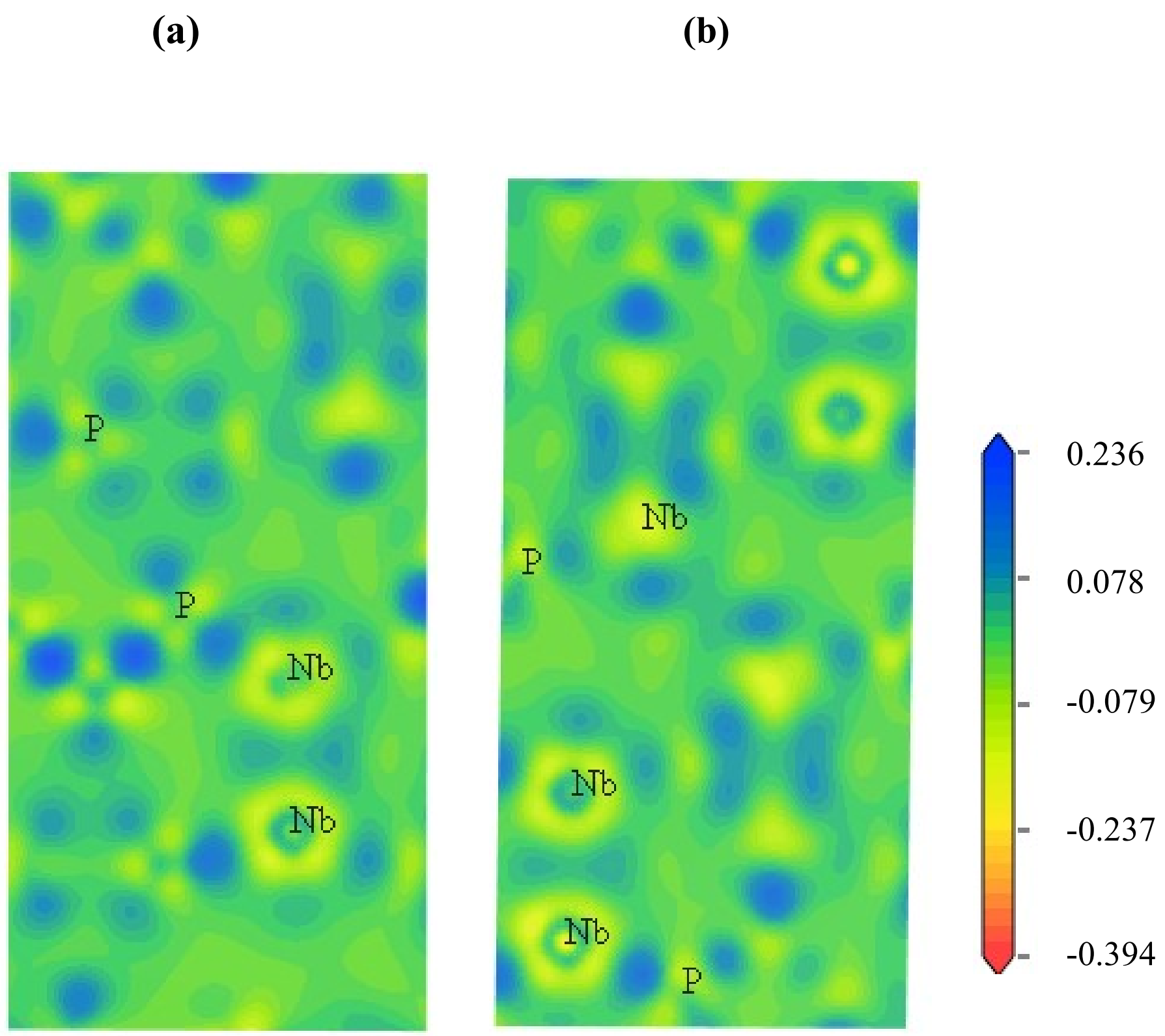

**Figure 5.** The electron density difference map of $Nb_2P_5$ in (a) (001) and (b) (111) plane.

## *3.7 Bond population analysis*

Mulliken bond population analysis is a widely employed technique in quantum chemistry to understand the bonding nature in compounds [28]. The Mulliken charge is linked to the vibrational properties of the molecule and quantifies how the electronic structure changes under atomic displacement. It is also related to the dipole moment, polarizability, electronic structure, charge mobility in reactions and other related properties of molecular systems. The spilling parameter and atomic charges resulting from these calculations are tabulated in Table 11. The analysis shows that the total charge for Nb atoms is larger than that for P atoms which mainly comes from 4*d* states of Nb atom. From Table 11, it is observed that the charges of Nb atoms are negative (behave as anion) and those of P atoms are positive (behave as cation). The atomic charges of Nb in $Nb_2P_5$ are -0.13 and -0.10 electron, depending of the atomic site. Whereas the atomic charges of P in $Nb_2P_5$ are +0.09, +0.05, +0.02, +0.04 and +0.21 electron, depending on the location in the crystal. All these values are deviated from the formal charge expected for purely ionic state (Nb: +4 and P: -3, -2,-1, 0). Thus the electrons are partially transferred from P to Nb atoms. This deviation reflects the presence of covalent bonds between Nb and P with ionic contributions. So, we can conclude that the bonding behaviors are the combination of ionic and covalent bonds. So, the bonding behavior found from the Mulliken atomic charges and effective ionic valence analysis is supported by electron density maps discussed in the preceding section. One can propose two possible electron transfer paths; intra or inter the Nb and P atoms.

The detailed effective ionic valence is also represented in Table 11. The difference between the formal ionic charge and the Mulliken charge on the cation species [30] is defined as Effective valence. It is used to measure the degree of covalency and/or iconicity. A value of zero effective valence is associated with a perfect ionic bond, while values greater than zero indicate increasing level of covalency. Table 11 shows that the effective valences for Nb is higher compared to P in $Nb_2P_5$. This again implies the presence of both ionic and covalent bonds are present in $Nb_2P_5$.

Since early times, it was demonstrated that the Mulliken bond population analysis display extreme sensitivity to atomic basis set and due to this, sometimes it gives results in contradiction to chemical intuition. On the contrary, Hirshfeld population analysis (HPA) [101] gives more meaningful result because it does not require a reference to basis set or their respective location. Taking this into account, we have calculated Hirshfeld charge of $Nb_2P_5$ using the HPA. For $Nb_2P_5$, HPA shows opposite result compared to Mulliken charge except +0.03 for P. Hirshfeld analysis shows atomic charges of Nb are +0.20 and +0.21 electronic charge. Hirshfeld analysis also shows atomic charges of P are +0.03, -0.15, -0.07, -0.08 and -0.15 electronic charge. Hirshfeld charge predicts that electrons are transferred from Nb to P and P to P. Both the approaches predict the presence of a mixture of covalent and ionic bondings between Nb and P atoms. We have also calculated effective valences of $Nb_2P_5$ using the Hirshfeld charge. Therefore, HPA predicts lower level of covalency of $Nb_2P_5$compared to the result we get from the Mulliken charge analysis.

**Table 11**

Charge spilling parameter (%), orbital charges (electron), atomic Mulliken charges (electron), formal ionic charge, effective valence (electron) and Hirshfeld charge (electron) in $Nb_2P_5$.

| Compound | Charge spilling | Species | Mulliken atomic population | | | | Mulliken charge | Formal ionic charge | Effective valence | Hirshfeld charge | Effective valence |
|---|---|---|---|---|---|---|---|---|---|---|---|
| | | | *s* | *p* | *d* | Total | | | | | |
| $Nb_2P_5$ | 0.70 | Nb | 2.41 | 6.52 | 4.20 | 13.13 | -0.13 | +4 | 3.87 | 0.20 | 3.80 |
| | | | 2.44 | 6.50 | 4.16 | 13.10 | -0.10 | +4 | 3.90 | 0.21 | 3.79 |
| | | P | 1.64 | 3.27 | 0.00 | 4.91 | 0.09 | 0 | 0.09 | 0.03 | 0.03 |
| | | | 1.62 | 3.33 | 0.00 | 4.95 | 0.05 | -3 | 2.95 | -0.15 | 2.85 |

| | | | | | | | | | |
|---|---|---|---|---|---|---|---|---|---|
| | 1.62 | 3.36 | 0.00 | 4.98 | 0.02 | -1 | 0.98 | -0.07 | 0.93 |
| | 1.65 | 3.31 | 0.00 | 4.96 | 0.04 | -2 | 1.96 | -0.08 | 1.92 |
| | 1.63 | 3.35 | 0.00 | 4.98 | 0.02 | -3 | 2.98 | -0.15 | 2.83 |

### *3.8 Theoretical bond hardness*

The study of the behavior of a material under varying load is an important part to understand its application, especially using as an abrasive resistant phase and radiation tolerant material [102]. Hardness is the property of a material which measures the resistance to indenting, cutting, bending, scratching, abrasion, etc. Relative hardness of a material is also commonly used in industry.

A compound with higher bond population or electron density, smaller bond length, and greater degree of covalent bonding is generally harder. Though the bulk modulus and shear modulus measures resistance to volume change and shape deformation of a solid, they give little direct information regarding hardness [103]. Intrinsic and extrinsic are the two types of hardness. Intrinsic hardness is mainly defined by the bonding. Generally the hardness of a single crystal is considered as intrinsic and that of nanocrystalline and polycrystalline solids as extrinsic. The following expressions can be used to calculate the intrinsic bond hardness and total hardness of a compound with mixture of bonding types including metallic bonding [104, 105]:

$$H_v^\mu = \left[\prod^{\mu}\left\{740\left(P^\mu - P^{\mu'}\right)\left(v_b^\mu\right)^{-5/3}\right\}^{n^\mu}\right]^{1/\Sigma n^\mu} \quad (37)$$

and

$$H_v = \left[\prod^{\mu}\left(H_v^\mu\right)^{n^\mu}\right]^{1/\sum n^\mu} \quad (38)$$

where $P^\mu$ is the Mulliken bond overlap population of the $\mu$-type bond, $P^{\mu'} = n_{free}/V$ is the metallic population (with $n_{free} = \int_{E_P}^{E_F} N(E)dE$ = the number of free electrons, $E_P$ and $E_F$ are the energy at the pseudogap and at the Fermi level, respectively), $n^\mu$ is the number of $\mu$-type bond, $v_b^\mu$ is the bond volume of $\mu$-type bond calculated by using the equation $v_b^\mu = (d^\mu)^3/\sum_v[(d^\mu)^3 N_b^\mu]$, $H_v^\mu$ is the bond hardness of $\mu$-type bond and $H_\mu$ is the hardness of the compound. The constant 740 is the proportional coefficient fitted from hardness of diamond [104, 105].

The calculated overlap population, bond length, total number of each type of bonds, bond hardness and the theoretical hardness of $Nb_2P_5$ are listed in Table 12. Experimentally, a crystal's hardness is determined from the sum of resistance of each bond per unit area to the indenter [105, 106]. As mentioned previously, to evaluate bonding characteristics in a material, the Mulliken bond population is a useful tool to employ. It allows us to allocate the electrons in some fractional manner among the various parts of the bonds. The Mulliken bond populations of a crystal define the degree of overlap of the electron clouds between two bonding atoms. Bond order is the overlap population of electrons between the atoms, and this is a measurement of the strength of the covalent bond between atoms and the strength of the bonds per the unit volume. High and low values of overlap population indicate dominance of covalent and ionic bonds, respectively. Whereas the exact zero overlap population value indicates that the bonds are perfectly ionic. For $Nb_2P_5$, we have found both positive and negative values of overlap population. The states of bonding or anti-bonding nature of interactions between the atoms are identified from the positive (+) and negative (-) values of bond overlap population, respectively [107]. The bond overlap population values of $Nb_2P_5$ indicate the presence of both bonding-type and anti-bonding-type interactions. Table 12 shows that most bonds exhibit strong covalent character. In $Nb_2P_5$, only three P-P bonds are of antibonding nature with bond population of -0.04, -0.14 and -0.09 and all the other bonds (Nb-Nb, Nb-P and P-P) are in the bonding state. The overall bond population remains positive and keeps the structure of $Nb_2P_5$ stable.

Beside hardness, low degree of covalency predicts the high temperature oxidation resistance of a material. A bonding with low degree of freedom implies that the atoms involved will break their bonding and might escape outward more easily and form protective oxide scale on the subsurface when thermal energy is supplied.

Mulliken population analysis is also useful in determining the metallic nature of bonds in a compound. Metallic bonding is soft and delocalized in nature and not directly related to the hardness of a material [108]. The calculated number of free electrons is $n_{free} = 4.76$. Metallic population of Nb-Nb, Nb-P, P-P bonds are very low and have the same value of 0.011. The metallicity of a crystal can be defined as,

$$f_m = P^{\mu'} / P^{\mu} \quad (39)$$

where, $P^{\mu'}$ and $P^{\mu}$ are metallic population and the Mulliken (overlap) population, respectively. The calculated results are listed in Table 12. The calculated values of hardness $H_v^{\mu}$ of a μ-type bond and total hardness $H_v$ of $Nb_2P_5$ are all disclosed in Table 12.

**Table 12**

The calculated Mulliken bond overlap population of μ-type bond $P^{\mu}$, bond length $d^{\mu}$(Å), metallic population $P^{\mu'}$, metallicity $f_m$, total number of bond $N^{\mu}$, bond volume $v_b^{\mu}$ (Å$^3$), hardness of μ-type bond $H_v^{\mu}$ (GPa)and Vickers hardness of the compound, $H_v$ (GPa) of $Nb_2P_5$.

| Compound | Bond | | $P^{\mu}$ | $d^{\mu}$ | $P^{\mu'}$ | $f_m$ | $N^{\mu}$ | $v_b^{\mu}$ | $H_v^{\mu}$ | $H_v$ |
|---|---|---|---|---|---|---|---|---|---|---|
| $Nb_2P_5$ | P1-P12<br>P7-P15<br>P5-P18<br>P3-P9 | P-P | 0.55 | 2.16 | 0.011 | 0.02 | 4 | 4.28 | 35.33 | |
| | P10-P18<br>P12-P16<br>P9-P19<br>P13-P15 | P-P | 0.52 | 2.20 | 0.011 | 0.02 | 4 | 4.53 | 30.36 | |
| | P3-P7<br>P1-P5 | P-P | 1.05 | 2.23 | 0.011 | 0.01 | 2 | 4.71 | 58.07 | |
| | P18-Nb8<br>P9-Nb2<br>P12-Nb4<br>P15-Nb6 | P-Nb | 0.97 | 2.50 | 0.011 | 0.01 | 4 | 6.64 | 30.23 | |
| | P 5-Nb5<br>P1-Nb1<br>P3-Nb3<br>P7-Nb7 | P-Nb | 0.45 | 2.53 | 0.011 | 0.02 | 4 | 6.88 | 13.04 | |
| | P19-Nb3<br>P16-Nb1<br>P10-Nb5<br>P13-Nb7 | P-Nb | 0.30 | 2.56 | 0.011 | 0.04 | 4 | 7.47 | 7.49 | |
| | P13-Nb3<br>P16-Nb5<br>P10-Nb1<br>P19-Nb7 | P-Nb | 0.74 | 2.60 | 0.011 | 0.01 | 4 | 7.56 | 18.51 | |
| | P13-P19<br>P10-P16 | P-P | -0.04 | 2.61 | 0.011 | -0.28 | 2 | 7.56 | -1.30 | |
| | P6-Nb7<br>P2-Nb3<br>P8-Nb5<br>P4-Nb1 | P-Nb | 0.82 | 2.61 | 0.011 | 0.01 | 4 | 7.56 | 20.54 | |
| | P14-Nb4 | P-Nb | 0.64 | 2.63 | 0.011 | 0.02 | 4 | 7.73 | 15.39 | 1.16 |

| | | | | | | | | |
|---|---|---|---|---|---|---|---|---|
| P17-Nb6 | | | | | | | | |
| P11-Nb2 | | | | | | | | |
| P20-Nb8 | | | | | | | | |
| P8-Nb8 | P-Nb | 0.31 | 2.63 | 0.011 | 0.04 | 4 | 7.73 | 7.32 |
| P2-Nb2 | | | | | | | | |
| P6-Nb6 | | | | | | | | |
| P4-Nb4 | | | | | | | | |
| P11-Nb4 | P-Nb | 0.42 | 2.65 | 0.011 | 0.03 | 4 | 7.91 | 9.63 |
| P20-Nb6 | | | | | | | | |
| P17-Nb8 | | | | | | | | |
| P14-Nb2 | | | | | | | | |
| P4-Nb2 | P-Nb | 0.58 | 2.65 | 0.011 | 0.02 | 4 | 7.91 | 13.40 |
| P8-Nb6 | | | | | | | | |
| P2-Nb4 | | | | | | | | |
| P6-Nb8 | | | | | | | | |
| P20-Nb7 | P-Nb | 0.66 | 2.68 | 0.011 | 0.02 | 4 | 8.18 | 14.45 |
| P17-Nb5 | | | | | | | | |
| P11-Nb1 | | | | | | | | |
| P14-Nb3 | | | | | | | | |
| P2-P14 | P-P | -0.14 | 2.79 | 0.011 | 0.08 | 4 | 9.23 | -2.75 |
| P8-P17 | | | | | | | | |
| P4-P11 | | | | | | | | |
| P6-P20 | | | | | | | | |
| P8-P20 | P-P | -0.09 | 2.97 | 0.011 | -0.12 | 4 | 11.14 | -1.34 |
| P4-P14 | | | | | | | | |
| P6-P17 | | | | | | | | |
| P2-P11 | | | | | | | | |

## *3.9 Optical properties*

The optical properties of a substance are defined by the interaction of photons (or incident electromagnetic wave) with that particular material (actually with the charge carriers in that material) when they are incident on the surface of that material. In recent decades, interest on optical properties of materials has increased many folds in materials science because of their close relations to the applications in integrated optics such as optical modulation, optoelectronics, optical information processing and communications, display devices and optical data storage. The energy or frequency dependent optical properties also provide with an important tool for studying energy band structure, state of impurity levels, excitons, localized defects, lattice vibrations, and certain magnetic excitations [109]. Electric and magnetic fields may lead to electric dipoles and magnetic moments, polarization charges, and induced current in solids. Clearly the electric and magnetic fields will not be uniform within a crystalline material but fluctuate from point to point reflecting the periodicity of the atomic lattice.

Since structural properties of $Nb_2P_5$show anisotropic nature, optical constants might also show direction dependency. Therefore, we have investigated optical properties of $Nb_2P_5$ for two different polarization directions [100] and [001] of the incident electric field. The electronic band structures properties shows that $Nb_2P_5$ is metallic in nature. The Drude damping correction is required for metallic materials [110, 111]. The calculation of the optical constants of $Nb_2P_5$ has done using a screened plasma energy of 0 eV and a Drude damping of 0.05 eV as prescribed in the CASTEP.

The optical properties of $Nb_2P_5$ was calculated by the frequency dependent dielectric function,

$$\varepsilon(\omega) = \varepsilon_1(\omega) + i\varepsilon_2(\omega) \tag{40}$$

The real part of the dielectric constant is associated with the degree of polarization and the velocity of electromagnetic wave inside the solid and the imaginary part gives an idea regarding how a compound absorbs energy from the incident electromagnetic field due to the frequency dependent oscillations of dipoles. In the condensed matter system, intra-band and inter-band electronic transitions contributes to $\varepsilon(\omega)$. Intra-band transitions play an important role at low energy, whereas the inter-band term depends strongly upon the details of the electronic band structure [112]. Other energy (frequency) dependent optical functions, such as the refractive index $n(\omega)$, extinction coefficient $k(\omega)$, optical reflectivity $R(\omega)$, absorption coefficient $\alpha(\omega)$, energy-loss function $L(\omega)$, and complex optical conductivity $\sigma(\omega)$ can be calculated from complex dielectric function $\varepsilon(\omega)$ of a material. The calculated optical functions of $Nb_2P_5$ are presented in Figure 6, for the photon energy range up to 25 eV for [001] and [100] polarization directions of the electric field.

Figure 6(a) illustrates the variation of real and imaginary parts of dielectric function, $\varepsilon(\omega)$ with photon energy of $Nb_2P_5$. It is observed from Fig. 6(a) that the values of $\varepsilon_2$ become zero at 20.6 eV indicating that the material will become transparent above 20.6 eV. Generally $\varepsilon_2$ becomes nonzero when the absorption occurs. Moreover, it shows that the effective plasma oscillations occur at ~20.91 eV, for both directions of polarization. This also locates the plasma frequency for $Nb_2P_5$. The real part of the dielectric constant shows conventional metallic character with Drude peak at low energy.

The refractive index of $Nb_2P_5$ is calculated using following equation: $N(\omega) = n(\omega) + ik(\omega)$, where $k(\omega)$ is extinction coefficient. The real part of the refractive index determines the phase velocity of the electromagnetic wave inside the sample, while the extinction coefficient (imaginary part) spectrum explains the amount of attenuation of the incident electromagnetic radiations while propagating through the compound. The complex refractive index is a crucial parameter for designing photoelectric device. The refractive index is one of the fundamental properties of a material, because it is closely related to the electronic polarizability of ions and the local field inside the material. The frequency dependence of the refractive index and the extinction coefficient of $Nb_2P_5$ for both polarization directions are displayed in Fig. 6(b). The static refractive index of $Nb_2P_5$ is quite high (~8) which shows this material has potential to be used in display device applications.

The optical conductivity of a material can be defined as the conduction of free charge carriers over a defined range of the photon energies. This is a dynamic response of mobile charge carriers which includes the photon generated electron hole pairs in semiconductors. Fig. 6(c) illustrates the calculated frequency dependent optical conductivity σ(ω) of $Nb_2P_5$ as a function of photon energy. For both polarization directions, photoconductivity starts with zero photon energy, which indicates that the materials has no band gap agreeing with the band structure (Fig. 2) and TDOS calculations (Fig. 3). The photoconductivity of both directions increases with photon energy, reaches to maximum (at 4.6 eV), decreases gradually with further increase in energy and tends to zero at around 20 eV. $Nb_2P_5$ has slightly higher conductivity around 4.8 eV along [001] direction compared to [100] direction, indicating a small anisotropy in optical properties. The optical conductivity of a system can be related to its imaginary part of the dielectric function. This is obvious from qualitative agreement in the response spectra of σ(ω) and $\varepsilon_2(\omega)$ in ultraviolet energies.

Figure 6(d) represents the reflectivity spectra of $Nb_2P_5$ as a function of photon energy. The maximum reflectivity value is about 80% and it occurs between the energy ranges 17 and 20 eV. R(ω) of $Nb_2P_5$ shows almost nonselective behavior over a broad energy range encompassing infrared, visible and mid ultraviolet region. The reflectivity spectra stay above 50% from 0 eV to 18 eV. This implies that $Nb_2P_5$ has a very good prospect to be used as an efficient reflector of radiation over a broad spectral range.

To understand a material's electrical nature, whether it is metallic, semiconducting or insulating, optical absorption coefficient is an important parameter. It also helps us to understand the optimum solar energy conversion efficiency of a material and how far light of specific energy can penetrate into the material before being absorbed. Optical absorption spectra also help to determine whether indirect or direct optical transition is occurring across the band gap [113]. Fig. 6(e) shows the absorption coefficient spectra of $Nb_2P_5$ along [100] and [001] polarization directions. It is seen from the figure that the optical absorption starts from zero photon energy. This is a hallmark of metallic band structure. Fig. 6(e) also shows that the absorption coefficient of $Nb_2P_5$ is quite high in the energy region from ~4 eV to ~16 eV. The position of peak value of absorption coefficient for [001] polarization is different from that [100] polarization. The peak value for [001] is around 6.6 eV, whereas for [100] the peak value is around 10.5 eV. The peak value for [001] is higher than that for [001] polarization direction. For both polarization directions, α(ω) decreases sharply at ~ 18 eV which in close agreement with the position of the loss peak and onset of collective charge (plasma) oscillation.

Fig. 6(f) shows the energy/frequency dependent electron energy loss function L(ω) for $Nb_2P_5$. The energy loss function is a parameter describing the energy loss of a fast electron traversing in a material. A material's loss function, absorption and reflection characteristics are closely interrelated. The highest peak in L(ω) represents the plasma resonance due to collective charge excitations at the corresponding plasma frequency $\omega_P$ of the material. In addition, the peaks of L(ω) also correspond to the trailing edges in the reflection spectra [114, 115], for instance, the peak of L(ω) is at about 20 eV corresponding the abrupt reduction of R(ω). Moreover, the peak in the loss function appears at $\varepsilon_2 < 1$ and $\varepsilon_1 = 0$ [27, 76, 116]. Energy loss function study is also helpful for understanding the screened charge excitation spectra, particularly the collective excitations produced by a swift electron traversing the solid. The incident light frequency (energy) at which the peak in L(ω) spectra occurs is known as the bulk screened plasma frequency. It is observed that for [100] and [001] polarizations, the peaks of L(ω) are located at 20.81 eV and 20.51 eV, respectively. The sharp loss peak represents the abrupt reduction in reflectivity and absorption coefficient of $Nb_2P_5$ (see Figs. 6(d) & (e), respectively).

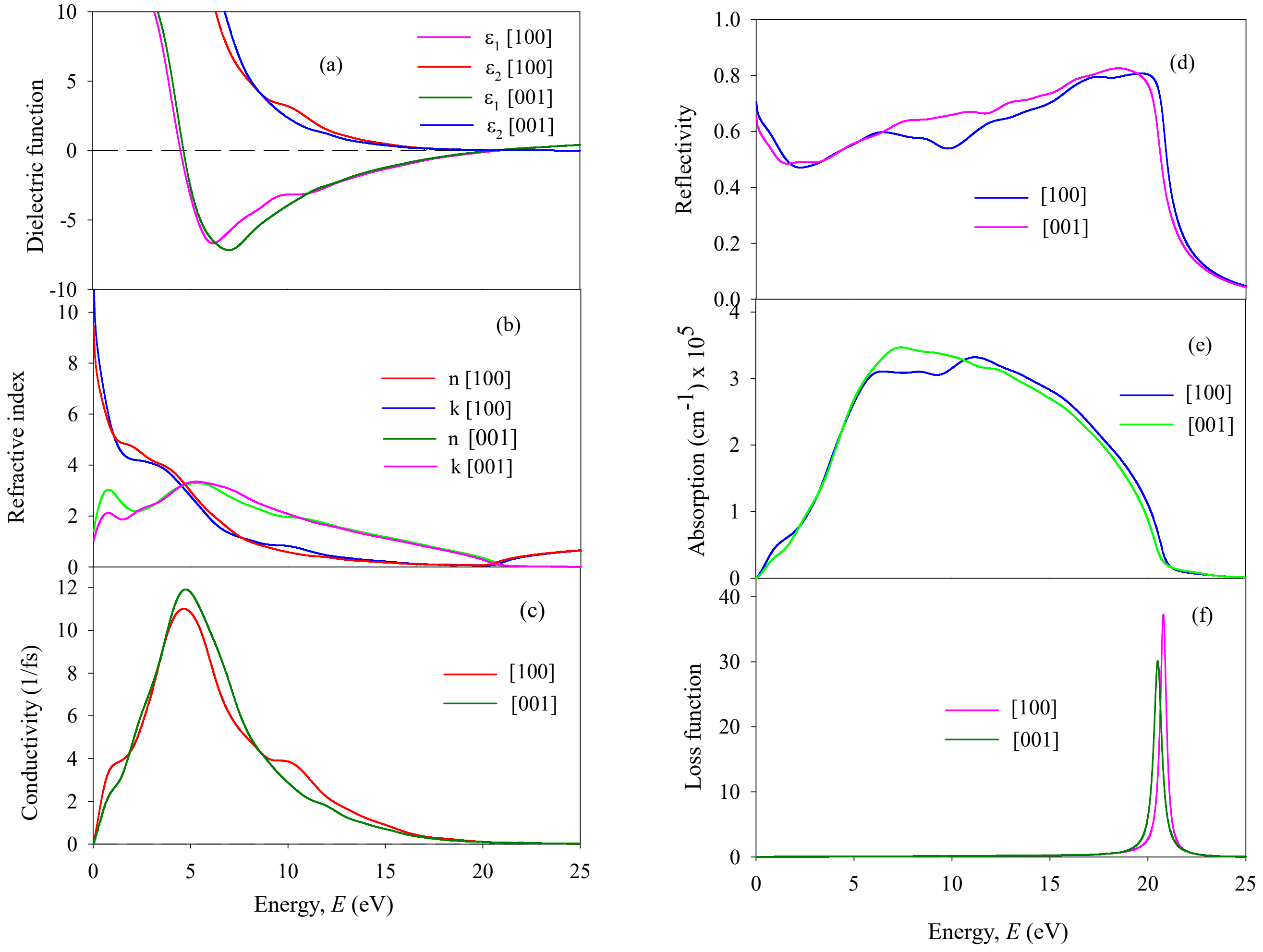

**Figure 6.** (a) $\varepsilon_1(\omega)$ and $\varepsilon_2(\omega)$, (b) $n(\omega)$ and $k(\omega)$, (c) σ(ω), (d) $R(\omega)$, (e) α(ω) and (f) $L(\omega)$ of $Nb_2P_5$, as a function of energy of incident photons.

## 4. Discussion and conclusions

$Nb_2P_5$ belongs to technologically important class of metallic phosphide materials. In this study we have investigated a large number of hitherto unexplored elastic, electronic, thermophysical, bonding and optoelectronic properties in details of $Nb_2P_5$. $Nb_2P_5$ possesses significant elastic anisotropy. The compound machinable and is marginally brittle in nature. The bonding character is mixed with notable covalent and ionic contributions. The band structure calculations reveal metallic characteristics with a large value of Coulomb pseudopotential which should contribute to the reduction of superconducting transition temperature [117]. The tendency of the formation of covalent bonding between the atomic species in $Nb_2P_5$ is also supported by the PDOS feature, where significant hybridization between Nb 4*d* and P 3*d* electronic orbitals are found. Covalent bonding with directional character tends to make the compound brittle. Mechanical strength in $Nb_2P_5$ is mainly dominated by bond bending contribution. Both the charge density distribution and electron density difference show moderately direction dependence. Comparatively higher value of the universal log-Euclidean index ($A^L$) implies that the bonding strength in $Nb_2P_5$ is anisotropic along different directions within the crystal and the compound may show layered characteristics. The hardness of $Nb_2P_5$ is moderate.

The calculated values of elastic moduli, Debye temperature, minimum thermal conductivity, melting temperature and thermal expansion coefficient imply that $Nb_2P_5$ has significant promise to be used as a thermal barrier coating (TBC) material.

The optical parameters are explored in details. $Nb_2P_5$ possesses high reflectivity and absorption coefficient over extended range of energy. The material also has a large static refractive index. All these features augur well for its use in the optoelectronic device sector.

Although the compound under study is elastically and mechanically anisotropic, the optical anisotropy is found to be quite low.

To summarize, we have studied the elastic, mechanical, bonding, acoustic, thermal, bulk electronic (band, DOS, charge density distribution, electron density difference) and optoelectronic properties of $Nb_2P_5$ in details in this paper. The elastic, electronic, bonding, acoustic, thermal and optoelectronic properties are studied in depth for the first time. The compound possesses several attractive mechanical, thermal and optoelectronic features which are suitable for engineering and device applications. We hope that the results obtained herein will stimulate researchers to investigate this binary metal phosphide in greater details in future, both theoretically and experimentally.

**Acknowledgements**
S. H. N. acknowledges the research grant (1151/5/52/RU/Science-07/19-20) from the Faculty of Science, University of Rajshahi, Bangladesh, which partly supported this work.

**Data availability**

The data sets generated and/or analyzed in this study are available from the corresponding author on reasonable request.

**Author Contributions**

M. I. N. performed the theoretical calculations, contributed in the analysis and wrote the draft manuscript. S. H. N. designed and supervised the project, analyzed the results and finalized the manuscript. Both the authors reviewed the manuscript.

**Additional Information**

**Competing Interests**

The authors declare no competing interests.